\documentclass[preprint,11pt]{elsarticle}



\usepackage[english]{babel}

\usepackage[letterpaper,top=2cm,bottom=2cm,left=3cm,right=3cm,marginparwidth=1.75cm]{geometry}

\usepackage{amsmath, amssymb}
\usepackage{graphicx}
\usepackage{subcaption}
\usepackage[colorlinks=true, allcolors=blue]{hyperref}
\usepackage{adjustbox}
\usepackage{float}
\usepackage{tikz}
\usetikzlibrary{arrows.meta}
\usepackage{caption}
\usepackage{orcidlink}
\usepackage{svg}
\usepackage{soul}
\usepackage{xcolor}
\usepackage{comment}
\usepackage{units}
\usepackage{ulem}
\usepackage{lineno}
\usepackage{soul}
\usepackage{bm}
\usepackage{cancel}
\usepackage{lineno}
\usepackage{etoolbox}
\usepackage{tikz}
\usetikzlibrary{arrows.meta, shapes.geometric}

\DeclareMathAccent{\svec}{\mathord}{letters}{126}

\usepackage{titlesec}
\titleformat{\section}
  {\normalfont\huge\bfseries}{\thesection}{1em}{}
\titleformat{\subsection}
  {\normalfont\large\bfseries}{\thesubsection}{1em}{}
\titleformat{\subsubsection}
  {\normalfont\normalsize\bfseries}{\thesubsubsection}{1em}{}
  \titleformat{\paragraph}
  {\normalfont\normalsize\bfseries}{\theparagraph}{1em}{}

\journal{Sound and Vibration}

\begin{document}

\begin{frontmatter}

\title{Vibroacoustic Underwater Noise from Fixed and Floating Offshore Wind Turbines}
%Operational Underwater Noise from Offshore Wind Turbines: A Comparative Vibroacoustic Analysis of Monopile and Floating Support Structures}

\author[1]{Raúl Sanz-Ramírez\corref{cor1}}
\ead{raul.sanz.ramirez@upm.es}
\author[1]{Martín de Frutos}
\author[1,3]{Guillen Campaña-Alonso}
\author[3]{Beatriz Méndez-López}
\author[1,2]{Esteban Ferrer}

\cortext[cor1]{Corresponding author}

\address[1]{ETSIAE-UPM, School of Aeronautics, Universidad Politécnica de Madrid, Plaza Cardenal Cisneros 3, 28040 Madrid, Spain}
\address[2]{Center for Computational Simulation, Universidad Politécnica de Madrid, Campus de Montegancedo, 28660 Boadilla del Monte, Spain}

\address[3]{Wind Energy Department, Centro Nacional de Energías Renovables (CENER), Ciudad de la Innovación, 7, 31621, Sarriguren, Spain}

\begin{keyword}
Offshore Wind turbines \sep Vibro-Acoustics \sep Underwater noise \sep Structure-borne noise \sep Floating wind turbines \sep Monopile foundations \sep Environmental acoustics
\end{keyword}

\begin{abstract}
Anthropogenic underwater noise from offshore wind turbines is a growing environmental concern, particularly in the context of large-scale deployment of bottom-fixed and floating devices.
    
This study presents a physics-based vibroacoustic framework to predict operational underwater noise emissions from offshore wind turbines and compares monopile-supported and floating configurations for a 10 MW turbine. The methodology combines time-domain aero-hydro-servo-elastic simulations with a frequency-domain acoustic formulation based on equivalent dipole sources and Green's function solutions that can account for the  underwater confinement between the free-surface and the seabed via the method of images.

The results show that floating configurations exhibit enhanced low-frequency acoustic emissions, producing up to 15\% ($\sim$ 15 dB) higher OASPL levels than monopile structures under equivalent water depths for frequencies below 10 Hz, as a result of additional rigid-body motions, while monopile structures tend to radiate more efficiently at higher frequencies associated with drivetrain excitations. Significant differences in the spatial distribution and directivity of the radiated sound field are also observed, with floating platforms displaying more complex three-dimensional radiation patterns and stronger direction-dependent variations, reaching approximately 20--25 dB in the 100--1000 Hz band, compared with the smoother and nearly axisymmetric response of the monopile configuration. The depth of the water strongly influences the propagation regimes and the overall sound levels, with shallow-water floating configurations showing variations of up to 7\% ($\sim$ 9 dB) in OASPL with respect to deep-water floating configurations.

The proposed framework enables the quantification of vibro-acoustic noise and provides a predictive tool for assessing underwater acoustic impacts during the design phase, thereby supporting environmentally informed offshore wind turbine design, as well as future regulatory and monitoring strategies.

%% LOW FREQUENCY SPL MISMTACH TAKEN FROM MAX IN SHALLOW AND MONOPILE SPL < 10 Hz

%% ENVIRONMENTAL % FROM SHALLOW TO DEEP TAKEN FROM MAXIMUM IN DISTANCE DECAY (~60m)

\end{abstract}

\end{frontmatter}
% -----------------------------
% Introducción
% \tableofcontents

\section{Introduction}
\label{sec: introduction}

Offshore wind energy plays a central role in the global energy transition and in the development of climate-neutral energy systems \cite{IRENA2019}. As the sector expands, offshore wind farms are moving towards larger turbine capacities, increased rotor diameters, and deployment in more energetic and deeper-water environments. These developments have led to the coexistence of different support concepts, ranging from conventional bottom-fixed foundations, such as monopiles, to emerging floating offshore wind turbine technologies.

Anthropogenic underwater noise is increasingly recognised as a significant environmental stressor in marine ecosystems, affecting organisms by masking biologically relevant signals, behavioural disturbances, and physiological stress \cite{Hildebrand2009,Slabbekoorn2010}. Although impulsive sources such as pile driving have been extensively studied and are subject to mitigation and regulatory frameworks, continuous low-frequency (0 - 1 kHz), noise from offshore infrastructure during operation remains comparatively less understood \cite{Tougaard2020}. As offshore renewables expand, the operational noise from wind turbines, which can travel over long distances underwater, can become a persistent and spatially widespread contributor to the underwater soundscape.

Operational underwater noise from offshore wind turbines originates from two primary physical pathways: airborne aeroacoustic generation at the rotor and structure-borne vibroacoustic radiation through the support structure. Aeroacoustic noise is generated by unsteady aerodynamic loading mechanisms such as turbulent boundary-layer–trailing-edge interactions, blade–wake interaction, and inflow turbulence. The potential for this noise to propagate into the water column has often been neglected due to the strong acoustic impedance contrast between air and water, which severely limits direct transmission across the interface. Nevertheless, our recent work \cite{AuthorYear_Aero} has shown that part of the aeroacoustic energy can propagate into the water column through airborne transmission, including air–water coupling, producing a broadband acoustic footprint underwater. However, this pathway is fundamentally distinct from the structure-borne transmission associated with tower vibrations, which can also radiate sound into the water. Disentangling these two mechanisms is necessary to correctly attribute the sources of underwater noise and propose palliative measures. %from offshore wind turbines and forms the focus of the present study.

The vibrational noise of offshore wind turbines arises from fluctuating aerodynamic thrust and torque, the excitation of the drivetrain, and hydrodynamic loading, which induce structural vibrations that propagate along the tower and the support structure before radiating into the surrounding water. The efficiency of this transmission and radiation process is strongly dependent on structural boundary conditions and on the typology of the foundation or the platform.
On the one hand, monopile-supported offshore wind turbines, currently among the most widely deployed configurations in shallow and intermediate water depths, provide a relatively stiff structural pathway through the tower, transition piece, and monopile foundation. In these systems, aerodynamic, drivetrain-induced, and hydrodynamic loads can be transmitted directly towards the seabed--structure interface and radiated into the water column through the vibration of the submerged support system. On the other hand, floating offshore wind turbines are designed for deeper waters, where fixed bottom foundations become technically or economically challenging. Their dynamics is governed not only by the aero-servo-elastic response of the rotor--nacelle-tower assembly, but also by rigid-body motions of the platform, hydro-elastic coupling, and mooring-line interactions \cite{Jonkman2009}. Compared with monopile-supported turbines, these additional low-frequency and coupled dynamic mechanisms may alter both the spectral content and the radiation efficiency of underwater noise emitted \cite{Tougaard2020}. 

Understanding these differences requires a unified vibroacoustic treatment capable of capturing the coupled aeroelastic--structural-acoustic mechanisms governing operational emissions, which is essential for assessing the underwater acoustic impact of both current and future offshore wind technologies. Despite growing recognition of these mechanisms, predictive design tools remain limited, and current environmental assessments frequently rely on post hoc measurements or simplified empirical scaling approaches that do not allow operators to evaluate underwater acoustic impacts during the early design phase. Therefore, there is a need for physics-based predictive tools capable of quantifying underwater noise as a function of turbine size, foundation type, and farm layout, enabling developers to optimise spatial configuration, turbine selection, and structural detailing with explicit consideration of the acoustic footprint. Beyond project-level optimisation, validated predictive frameworks can also inform the development of science-based regulatory thresholds and monitoring protocols by linking measurable structural response parameters to radiated sound levels. In this way, advances in aeroacoustic and vibroacoustic modelling can provide traceable metrics for standardisation, compliance verification, and long-term monitoring, while contributing to more robust environmental regulation of offshore wind deployment.

%Despite these technological developments, a clear scientific gap persists: a systematic, physics-based vibroacoustic comparison between monopile-supported and floating offshore wind turbines under steady operational conditions is currently lacking. Existing research predominantly addresses construction-phase noise or relies on simplified empirical models for operational sound emissions \cite{Bailey2010}. The coupled aeroelastic--hydrodynamic--structural pathways governing structure-borne underwater noise radiation remain insufficiently resolved, particularly for floating systems.

In summary, the objective of this study is to develop predictive vibroacoustic tools capable of quantifying underwater noise emissions from offshore wind turbines in the operational regime and to apply these tools to a controlled comparison between fixed (monopile) and floating configurations subjected to equivalent aerodynamic loading conditions. The analysis focuses on low-frequency structure-borne noise, identifies dominant transmission mechanisms, and evaluates differences in acoustic radiation efficiency associated with foundation typology.
The specific scientific contribution of this work is threefold: (i) formulation of an integrated modelling framework coupling aeroelastic loading, structural dynamics, fluid--structure interaction, and underwater acoustic propagation; (ii) definition of measurable acoustic descriptors enabling consistent comparison between support structures; and (iii) identification of physical mechanisms responsible for differences in operational noise signatures between bottom-fixed and floating offshore wind turbines. By providing a quantitative and mechanistic assessment of vibroacoustic behaviour across turbine concepts, this study establishes a foundation for environmentally informed offshore wind design and future mitigation strategies.

\subsection{State of the art}
\label{subsec: state of the art}

Early studies established that offshore wind turbine operational noise is generally dominated by low-frequency components and tonal features associated with mechanical and structural excitation, with sound transmitted from the nacelle and drivetrain through the tower and support structure into the water column \cite{Madsen2006,Tougaard2009}. Measurements of different turbine and foundation types showed that operational noise can be audible to marine mammals over distances ranging from tens of metres to several kilometres, depending on species hearing sensitivity, ambient noise, propagation conditions and turbine operation \cite{Tougaard2009}. Although construction noise, particularly pile driving, has historically received most regulatory attention, operational noise is continuous throughout the lifetime of a wind farm and may therefore contribute persistently to the underwater soundscape.

For bottom-fixed turbines, the most detailed field evidence has been obtained from monopile-supported machines. Pangerc et al.~\cite{Pangerc2016} measured underwater sound in close proximity to two 3.6 MW monopile turbines and reported tonal components, with most of the acoustic energy below approximately 420 Hz. Their measurements showed that the radiated sound field depends on turbine operating conditions and contains dominant one-third-octave-band contributions around 160 Hz, highlighting the importance of structural transmission paths in operational underwater noise. Subsequent synthesis work by Tougaard et al.~\cite{Tougaard2020} reviewed available measurements from operating offshore wind turbines and concluded that, although operational turbine noise is typically lower than ship noise in the same frequency range, cumulative contributions from large wind farms can elevate low-frequency sound levels over distances of several kilometres under low ambient-noise conditions. St\"ober and Thomsen~\cite{Stober2021} further extrapolated the available measurements to larger future turbines and suggested that increasing turbine power capacity may increase operational source levels, implying that the acoustic relevance of operational noise may grow as offshore wind farms scale towards larger machines.

Recent measurements have reinforced the connection between turbine operation, structural excitation, and underwater sound. Yoon et al.~\cite{Yoon2023} measured operational underwater noise from a 3 MW offshore wind turbine in Korea and showed that tonal frequencies varied with rotor speed, with a pronounced increase in sound level around 198 Hz under rated operating conditions. These results support the view that operational underwater noise is not only a function of distance and bathymetry, but also of the aero-servo-mechanical state of the turbine. More recent reviews and meta-analyses have also identified propagation distance, wind speed, turbine power, drivetrain technology, and foundation type as relevant variables controlling measured sound pressure levels \cite{Ge2025}. In particular, monopile foundations are often associated with efficient transmission of structure-borne vibrations into the surrounding water because the tower--transition-piece--monopile system provides a comparatively direct and stiff vibro-acoustic pathway.

In parallel, government-commissioned and agency-funded studies have contributed significantly to the operational-noise evidence base. In the United Kingdom, early COWRIE-funded work measured and interpreted underwater sound during both construction and operation of offshore wind farms in UK waters \cite{Nedwell2007}. In Scotland, Marmo et al.~\cite{Marmo2013}, in a Scottish Government report, modelled the noise effects of operational offshore wind turbines considering different foundation configurations, including jacket, monopile, and gravity-base structures. This study is relevant because it treated operational underwater noise as a foundation-dependent transmission problem, rather than only as a far-field propagation problem. It therefore anticipated the need for models able to link turbine excitation, support-structure response, foundation typology, and underwater acoustic radiation. Similar government or agency-backed evidence has been generated in other jurisdictions, including BOEM-supported monitoring at the Block Island Wind Farm in the United States \cite{HDR2019,HDR2020,Amaral2019}, Belgian monitoring of operational sound from offshore wind farms in the North Sea \cite{Degraer2021,Norro2016}, Dutch TNO studies on underwater-noise measurement and monitoring for offshore wind licensing \cite{deJong2011,TNO2016}, and German cross-project evaluations of operational offshore wind turbine noise \cite{Bellmann2024}. Together, these studies show that operational underwater noise is increasingly recognised in regulatory and environmental-assessment practice, but also that most assessments remain measurement-based, empirically scaled, or propagation-based.

Compared with monopile-supported turbines, the literature on floating offshore wind turbine underwater noise is still emerging. Floating systems introduce additional dynamic mechanisms, including platform surge, sway, heave, roll, pitch, and yaw, as well as mooring-line dynamics and hydro-elastic coupling. These mechanisms can alter both the excitation transmitted through the support structure and the acoustic radiation pathways. In addition, floating devices cannot damp vibrations through seabed coupling and therefore exhibit a different pattern of acoustic emissions. Recent measurements and case studies from Hywind Scotland indicate that floating systems can generate not only continuous operational tonal sound, but also transient sounds associated with mooring and anchoring components, such as creaks, snaps, bangs, and rattles \cite{Burns2022,Pace2024}. This suggests that floating offshore wind turbines should not be treated as a simple extension of bottom-fixed systems: the platform, moorings, and dynamic cable may introduce additional low-frequency and transient vibro-acoustic sources.

Recent Scottish floating-wind studies have provided some of the first operational-noise datasets for commercial-scale floating offshore wind turbines. Risch et al.~\cite{Risch2023} characterised the underwater operational noise from two floating offshore wind farms deployed off the east coast of Scotland: Kincardine, based on semi-submersible foundations, and Hywind Scotland, based on spar-buoy foundations. Their analysis identified low-frequency tonal components associated with turbine operation, together with transient broadband signals attributed to mooring-system interactions. Similarly, the Hywind Scotland sound-source characterisation study by Burns et al.~\cite{Burns2022} highlighted the complexity of floating-turbine acoustic emissions, where drivetrain-related tonal noise can coexist with platform and mooring-related transient contributions. These studies provide important evidence that floating offshore wind noise involves additional source mechanisms that are not present or are much less significant in conventional monopile-supported turbines.

Because direct measurements from floating wind farms remain limited, several recent studies have relied on propagation modelling to evaluate potential impacts. Baldachini et al.~\cite{Baldachini2024,Baldachini2025} modelled underwater sound from proposed floating offshore wind farms in the Central Mediterranean using range-dependent propagation tools, including parabolic-equation and beam-tracing approaches. Their results show that the environmental relevance of floating wind turbine noise is strongly dependent on local bathymetry, sound-speed structure, source assumptions, and species-specific acoustic thresholds. In addition, project-level environmental-impact documentation, such as the Pentland Floating Offshore Wind Farm underwater-noise modelling and impact-assessment studies \cite{Pentland2022a,Pentland2022b}, illustrates the current regulatory practice in which prescribed or back-calculated source levels are combined with propagation models to estimate received levels and potential ecological effects.

From a modelling perspective, current offshore wind underwater-noise assessments often separate the problem into source description and propagation modelling. Once a source level is prescribed, either from measurements, empirical scaling laws, or conservative assumptions, propagation can be predicted using range-dependent transmission-loss models, including ray-based, normal-mode, wavenumber-integration, beam-tracing, or parabolic-equation approaches. These models can account for bathymetry, seabed properties, water-column sound-speed profiles, and frequency-dependent attenuation. However, the source term itself is often not predicted by first principles. In many cases, the turbine is represented as an equivalent underwater acoustic source, rather than deriving the radiated sound from the coupled aerodynamic, drivetrain, hydrodynamic, structural, and acoustic response of the system.

%More detailed physics-based formulations have been developed for related offshore foundation problems and provide a useful basis for operational vibro-acoustic modelling. Although much of this work was originally motivated by pile-driving noise, coupled pile--water--soil models identify the structural wave propagation, fluid loading, seabed coupling, and acoustic radiation mechanisms that are also relevant when operational loads excite the support structure. For example, Tsouvalas and Metrikine~\cite{Tsouvalas2013} developed a semi-analytical model in which the pile is represented as an elastic shell coupled to a compressible acoustic fluid and to the seabed through dynamic impedances. More recent work has begun to address underwater noise as a design-dependent vibro-acoustic problem. Zhou and Guo~\cite{Zhou2023} proposed a compact circular liner around the submerged part of an offshore wind turbine support structure and analysed its noise-reduction performance using equivalent-medium theory, transfer-matrix analysis, and finite-element acoustic--structure interaction simulations. This type of study is important because it moves beyond post hoc impact assessment and demonstrates how structural or acoustic design modifications may reduce underwater sound radiation at source.

Overall, existing studies provide useful first-order impact assessments and an increasingly broad empirical evidence base, but they also reveal a major gap in the state of the art: source models for offshore turbines, and especially for floating systems, remain uncertain because the relative contributions of drivetrain excitation, aerodynamic loading, platform vibration, mooring-line dynamics, and hydrodynamic loading are not yet fully quantified. Current propagation-based assessments can estimate far-field received levels once a source term is prescribed, but they generally do not derive that source term from the coupled aero-servo-hydro-elastic and vibro-acoustic response of the turbine. The present work addresses this gap by quantifying vibro-acoustic noise generation in monopile-supported and floating offshore wind turbines, with particular emphasis on the role of support-structure dynamics and hydrodynamic excitation in shaping the radiated underwater sound field.

The remainder of this work is organised as follows. Section~\ref{sec: methodology} describes the methodology, including the structural simulations, the acoustic modelling approach, and the underlying assumptions. Section~\ref{sec: results} presents the main results in terms of structural response and radiated underwater sound characteristics. Section~\ref{sec: discussion} discusses the implications of these findings for different offshore support structures, and Section~\ref{sec: conclusions} summarises the main conclusions. 

% Metodología
\section{Methodology}
\label{sec: methodology}

The present study combines structural dynamic simulations with frequency-domain acoustic modelling to estimate the underwater noise radiation of a 10 MW offshore wind turbine in two support configurations: monopile and floating platform. The overall methodology consists of three sequential stages. First, time domain aero-hydro-servo-elastic simulations are performed using OpenFAST to obtain the structural response of the turbine under rated operation conditions. Second, nodal accelerations are converted into equivalent dipole force sources based on an effective mass formulation. Finally, the acoustic pressure field is computed in the frequency domain by solving the linear Helmholtz equation using analytical Green's functions, including seabed and free-surface boundary conditions via the method of images. A detailed description of each step is provided below.

% ##################################################################### %
\subsection{Turbine and Support Structure Definition}
\label{subsec: structure definition}

\begin{table}[htbp]
\centering
\begin{tabular}{@{}lc@{}}
\hline
Characteristic & DTU~10~MW \\
\hline
Hub height [m] & 119.0 \\
Rotor diameter [m] & 178.4 \\
Nominal wind velocity [m/s] & 11.4 \\
Rotor angular velocity [rpm] & 9.6 \\
Blade tip velocity [m/s] & 90.0 \\
Blade Pitch [deg.] & 10.5 \\
\hline
\end{tabular}
\caption{Geometrical and operational (rated) conditions for the DTU 10 MW wind turbine\cite{bak2013dtu10mw}.} \label{tab:WT_nominal_cond}
\end{table}

We consider the DTU 10 MW \cite{bak2013dtu10mw} offshore wind turbine as a representative turbine of modern large-scale offshore systems.  
The DTU 10 MW Reference Wind Turbine can be employed as a baseline model for both fixed-bottom and floating offshore wind turbine configurations. It has been widely used in the literature as a reference rotor–nacelle assembly, decoupled from a specific support structure. For fixed-bottom applications, the turbine has been mounted on monopile foundations for structural design and load analysis studies \cite{bak2013dtu10mw,MonopileDTU10MW}. In parallel, it has also been widely adopted in floating offshore wind turbine research, where it is integrated with various platform concepts and used in both numerical and experimental investigations \cite{Madsen2020DTU10MW,Johannessen2018DTU10MW}. This flexibility makes the 10 MW DTU turbine a standard benchmark for comparative studies across different offshore support structures.

 The main characteristics, including nominal operating conditions, are detailed in table \ref{tab:WT_nominal_cond}. The turbine has a hub height of approximately 119 m and a rotor diameter of 178 m. 
In this work, all simulations are performed under nominal operating conditions that correspond to a steady uniform inflow wind speed of 11.4 m/s. Under these conditions, the turbine operates at a rotational speed of 9.6 rpm, resulting in a tip-speed ratio of $\lambda = 6$. This operating point ensures steady aerodynamic loading and allows analysis under statically stationary structural response conditions.
%
% The diameter of the tower varies from 8.3 m at the base to 5.5 m at the yaw bearing location. 
%\vspace{5mm}

\begin{figure}[H]
    \centering
    \includegraphics[trim = 0 100 0 100, clip =true, scale=0.3]{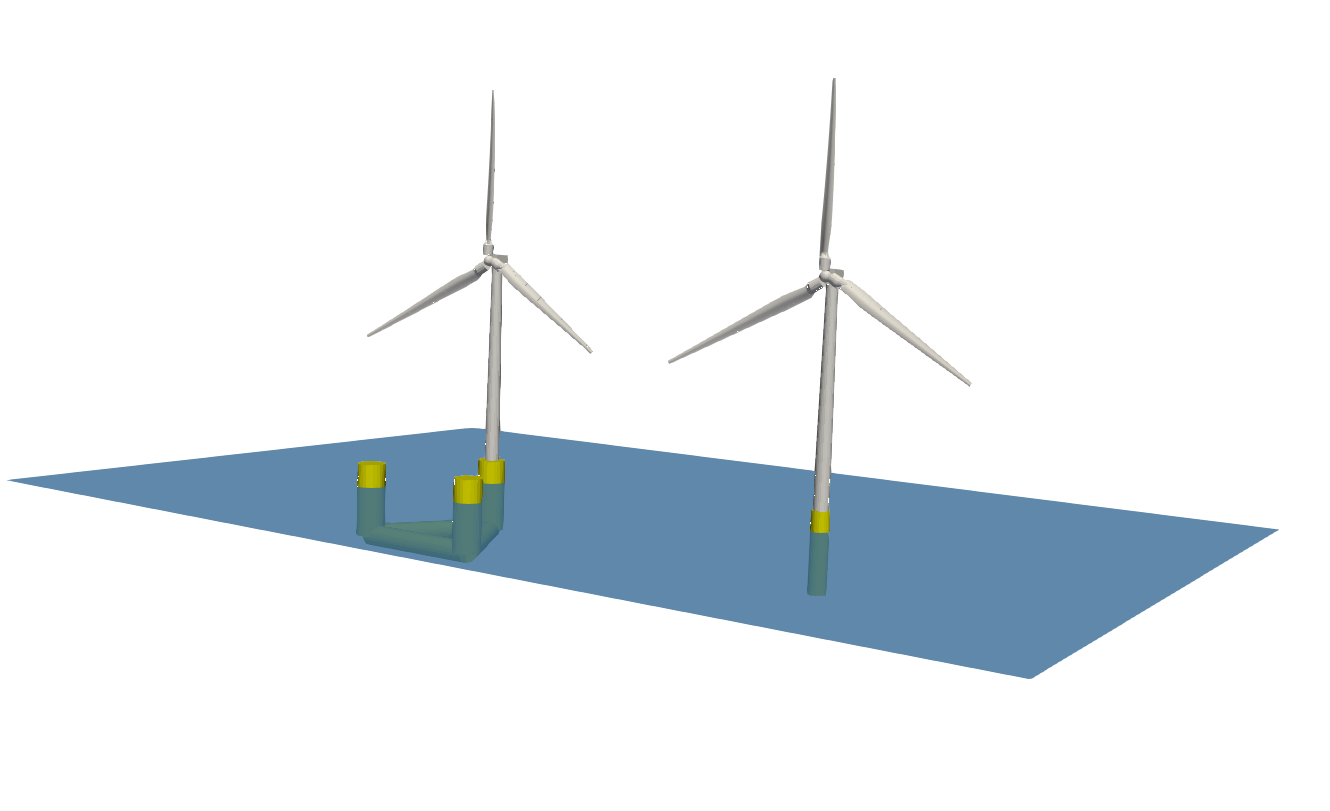}
    \caption{10 MW turbine configurations: Left, floating and Right, fixed monopile.}
    \label{fig: turbine configuration}
\end{figure}

In the following, we consider the 10 MW turbine in monopile and floating configurations. Figure \ref{fig: turbine configuration} illustrates both concepts, which are detailed here.

% ---------------------------------------------------------------------- %
\subsubsection{Monopile configuration}
The fixed bottom configuration \cite{bortolotti2019iea37} consists of a cylindrical monopile with a total length of 40 m and a constant diameter of 9 m with variable wall thickness ranging between 0.10 m and 0.155 m. The turbine is assumed to be installed at a site with a depth of 30 m in water, resulting in 30 m of submerged substructure and 10 m of elevation above the mean sea level. The tower base is therefore located 10 m above the water surface. The monopile provides a stiff boundary condition on the seabed, restricting global rigid-body motion and leading to structural dynamics dominated by bending modes of the tower-substructure assembly.

% ---------------------------------------------------------------------- %
\subsubsection{Floating configuration}
The floating configuration is based on a semi-submersible platform composed of three vertical cylindrical columns and three horizontal pontoons that connect their bases. This floating platform has been developed by CENER (the National Renewable Energy Centre of Spain) and is publicly available \cite{cener_innwind_10MW, innwind_D1_2_1, innwind_D4_3_3}.
Each column has a circular cross-section with a diameter of 14.5 m and a total height of 37.5 m. The columns extend 12 m above the mean water level, leaving approximately 25.5 m submerged. The columns are connected at their base by rectangular pontoons of 66 m length and cross-sectional dimensions of 7 m x 10.875 m.
The structural mass of the empty platform is approximately $1.76\times 10^6$ kg. To ensure hydrostatic stability, the platform is partially ballasted with seawater, resulting in a total operational mass of $2.36\times 10^7$ kg. The hydrostatic restoring stiffness in pitch is approximately $2.992\times 10^9$ N·m / rad, which provides the primary restoring mechanism against wave-induced rotational motions. The buoyancy centre is located at $(-38.11,0,-17.32)$ m, while the centre of gravity is located at $(-39.91,0,-18.63)$ m in the global reference frame, its origin located at the base of the tower.
Unlike the monopile case, the floating platform introduces additional low-frequency rigid-body motions, which significantly influence the global dynamic response and potentially the radiated acoustic field.
The geometric and mass properties described above define the structural domain from which vibration-induced acoustic radiation is computed.

% ##################################################################### %
\subsection{Structural Dynamics Simulations with OpenFAST}
\label{subsec: structural dynamics}

% ---------------------------------------------------------------------- %
\subsubsection{Governing equations}

The aerodynamic response, structural deformation, and fully coupled dynamics of wind turbines are calculated using OpenFAST \cite{openfast2023}. OpenFAST solves the coupled aero-hydro-servo-elastic equations of motion of the wind turbine system in the time domain. The resulting global structural dynamics can be expressed in reduced-order form as follows:
\begin{equation}
\mathbf{M\ddot{q}}(t) + \mathbf{C\dot{q}}(t) + \mathbf{Kq}(t) = \mathbf{F}_{ext}(t),
    \label{eq: OpenFAST dynamic equation}
\end{equation}
where $\mathbf{M}$ is the generalised mass matrix, $\mathbf{C}$ is the damping matrix, $\mathbf{K}$ is the stiffness matrix, $\mathbf{q}(t)$ represents the generalised coordinates, and $\mathbf{F}_{ext}(t)$ includes aerodynamic, hydrodynamic, gravitational, and control loads, which include arbitrary time series of external forces.

% ---------------------------------------------------------------------- %
\subsubsection{Structural modelling and Degrees of Freedom}
The tower and blades are modelled using beam theory formulations implemented in \texttt{ElastoDyn}, while the structural dynamics of the support structure are modelled using \texttt{SubDyn} for both the monopile and the floating configurations. \texttt{SubDyn} represents the substructure through a first-order finite elements formulation based on Timoshenko beam elements with distributed mass and stiffness properties \cite{branlard2020superelement}. This formulation enables the elastic response of the support structure to be resolved, including bending and axial deformations.

Hydrodynamic loads are computed in \texttt{HydroDyn} and transferred to \texttt{SubDyn} as external forces acting on discretised structural elements. \texttt{HydroDyn} accounts for wave excitation, added mass, radiation damping, and hydrostatic restoring effects. The coupled \texttt{SubDyn-HydroDyn} interaction therefore captures the elastic structural response under hydrodynamic loading in both configurations.\vspace{2mm}

The primary structural degrees of freedom include:
\begin{itemize}
    \item 2 Tower fore-aft and side-side bending modes.
    \item 2 Blade flap-wise and 1 edge-wise bending modes.
    \item 8 retained elastic bending modes of the substructure (via \texttt{SubDyn}).
\end{itemize}

In the floating configuration, additional low-frequency rigid-body degrees of freedom are introduced, corresponding to platform surge, sway, heave, roll, pitch, and yaw motions. These global platform modes typically lie in the very-low frequency range and significantly influence the dynamic response of the overall system. In contrast, the monopile configuration is bottom-fixed and, therefore, does not exhibit global rigid-body motions, resulting in a comparatively stiffer low-frequency response.

Due to the large structural dimensions and the beam-type modelling approach, the dominant eigenfrequencies of the whole system are located in the low-frequency range (generally below 5 Hz), corresponding to global bending modes and platform motions in the floating case. However, operational offshore wind turbines also exhibit higher-frequency vibrations associated with drivetrain components \cite{betke_noise_offshore,utgrunden2003underwater,Yoon2023,odegaard2020vvm}. To account for their contribution to acoustic radiation, higher-frequency harmonic excitations representative of drivetrain-induced vibrations \cite{Wang_2020_wind, Wang_2021} are introduced in a controlled manner at the hub location. %The virtual excitation introduced at the hub is described in detail in \ref{ap: drivetrain}. \textcolor{red}{yo el anexo B lo pondria aqui.. las excitaciones del drivetrain son importantes}

% ---------------------------------------------------------------------- %
\subsection{Drivetrain Excitation Model}
\label{subsec: drivetrain}

The force of the synthetic generator is constructed from the rated operating condition of the 10~MW turbine (9.6~rpm). Since the OpenFAST aero-hydro‑servo‑elastic code does not resolve the dynamics of the internal gearbox, an explicit set of tonal excitations is introduced in the nacelle to reproduce them. The frequencies follow the drivetrain kinematics reported by Wang~et~al.~\cite{Wang_2021} for a compact medium‑speed 10~MW gearbox, and include shaft rotation harmonics (LSS, IMS, HSS) as well as the fundamental gear‑mesh components of the three stages (GM1, GM2, GM3). This set captures the dominant narrowband spectral lines observed in gearbox vibration measurements.

% ---------------------------------------------------------------------------- %
\subsubsection{Amplitude scaling}

The amplitudes of the gear-mesh harmonics are assumed to follow an exponential decay with harmonic order $m$,
\begin{equation}
A_m = A_1\,\mathrm{e}^{-\beta (m-1)},
\label{eq:decay_b}
\end{equation}
with $\beta = 0.5$, which reproduces the rapid drop found in the experimental gear spectra \cite{Kahraman1994}. To reflect the physical hierarchy of the drivetrain, the base amplitudes $A_1$ are further weighted by shaft stage: the low‑speed shaft (driven by aerodynamic torque) receives the largest contribution, while energy progressively attenuates through the intermediate and high‑speed shafts \cite{Peeters2006,Nejad2016,Guo2012}. Gear-mesh amplitudes are ranked similarly, with the first stage (GM1) carrying the highest level.

% ---------------------------------------------------------------------------- %
\subsubsection{Modal structural amplification}

The drivetrain housing and shafts possess lightly damped torsional resonances that amplify the excitation near their natural frequencies. The modal transfer function used here is derived from classical vibration theory for a set of single‑degree‑of‑freedom oscillators \cite{Inman2014}. For each excitation frequency $f_k$, the amplification factor is
\begin{equation}
H(f_k) = \sqrt{\sum_n \frac{1}{\bigl[1-(f_k/f_n)^2\bigr]^2 + (2\zeta f_k/f_n)^2}},
\label{eq:H_b}
\end{equation}
where $f_n$ are the torsional natural frequencies obtained from the dynamic analysis of the 10~MW drivetrain \cite{Wang_2021} and $\zeta = 2\%$ is a typical structural damping ratio for steel components. The scaled amplitude is then applied
\[
A_{\text{tot}}(f_k) = A(f_k)\,H(f_k).
\]

% ---------------------------------------------------------------------------- %
\subsubsection{Normalisation and time‑series reconstruction}

The discrete excitation spectrum is first normalised to the unit root‑mean‑square value,
\begin{equation}
\sum_k \frac{A_k^2}{2} = 1,
\label{eq:rms_b}
\end{equation}
so that it can be later scaled to match the total RMS force level reported for the 10~MW drivetrain \cite{Wang_2020_wind}. A time‑domain representation is obtained by harmonic superposition,
\begin{equation}
F(t) = \sum_k A_k \sin(2\pi f_k t + \phi_k),
\label{eq:F_t_b}
\end{equation}
with random phases $\phi_k\sim\mathcal{U}[0,2\pi]$ that avoid artificial coherence between harmonics. For gear‑mesh components, a narrow Gaussian frequency cluster is centred at each nominal mesh line, acknowledging that real drivetrains exhibit small speed fluctuations and load ‐-dependent frequency smearing. This produces a realistic spectral bandwidth while preserving the tonal structure of the excitation.

\begin{figure}[htbp]
    \centering
    \includegraphics[trim = 0 0 0 24, clip=True, width=0.85\textwidth]{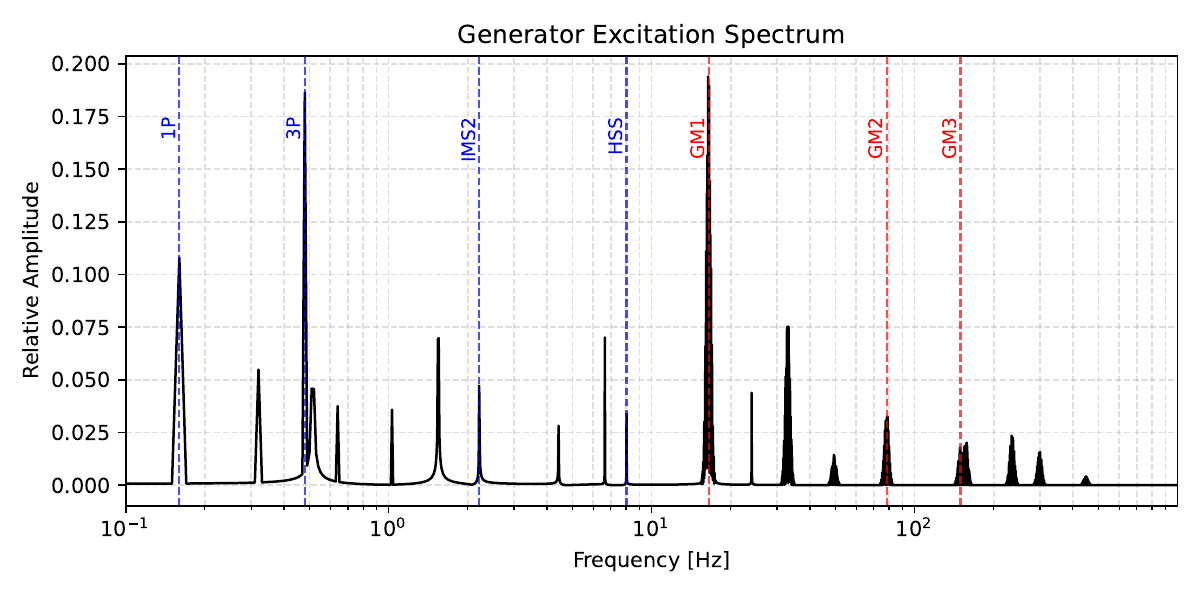}
    \caption{Frequency spectrum of the reconstructed generator excitation signal (normalised to unit RMS). Vertical lines indicate the nominal drivetrain excitation frequencies derived from shaft harmonics and gearbox mesh kinematics. 1P and 3P denote the first and third shaft rotational frequency harmonics. LSS, IMS and HSS denote the low‑, intermediate‑ and high‑speed shafts, respectively, while GM1–GM3 correspond to the fundamental gear mesh frequencies of the gearbox stages.}
    \label{fig: generator spectrum}
\end{figure}

% ##################################################################### %
\subsection{Acoustic Formulation}
\label{subsec: Acoustic Formulation}

The acoustic model employed in this work is based on the linearised compressible Euler equations with external forcing \cite{wagner1996wind}. Structural vibrations are represented as localised body forces acting on the surrounding fluid such that

\begin{equation}
F_i(\mathbf{x},\mathbf{y},t)=f_i(t)\delta(\mathbf{x}-\mathbf{y}),
\end{equation}
where $\mathbf{y}$ denotes the source position and $\delta$ is the Dirac delta distribution.

Assuming small perturbations about a quiescent mean state and isentropic compressibility, the resulting acoustic pressure field corresponding to a dipole source reads:

\begin{equation}
\tilde{p}'(\mathbf{r},\omega)
=
\frac{1}{4\pi r}
\left(
\frac{1}{r}-ik_0
\right)
\left(
\tilde{\mathbf{f}}(\omega)\cdot\hat{\mathbf{r}}
\right)
e^{ik_0 r},
\label{eq: dipole solution}
\end{equation}
where $\hat{\mathbf{r}}=\mathbf{r}/|\mathbf{r}|$. This formulation is detailed in \ref{ap: Acoustic Formulation} and reflects the classical dipole radiation mechanism associated with fluctuating structural forces acting on the surrounding fluid (without net mass injection) \cite{lighthill1952sound,wagner1996wind}. Under the small-amplitude vibration regime considered here, higher-order non-linear
and quadrupole volume-source contributions can be neglected. For compact sources with a low Mach number, these terms are higher order than the surface-loading dipole term, which therefore provides a leading-order description of the radiated acoustic field \cite{lighthill1952sound,Curle1955,FfowcsWilliamsHawkings1969}.

In the final discrete model, each node is modelled as a point dipole located at its global spatial coordinate. The dipole orientation is defined by the direction of the reconstructed force vector. The total acoustic pressure field is obtained by linear superposition of all dipole contributions:

\begin{equation}
    \tilde{p}(\mathbf{x},\omega) = \sum_{n=1}^{N_d}\frac{1}{4\pi r_n}\left(\frac{1}{r_n}-ik_0\right) \left(\tilde{\mathbf{f}}_n(\omega)\cdot \hat{\mathbf{r}}_n \right) e^{ik_0 r_n}
    \label{eq: linear superposition dipole solution}
\end{equation}
where $r_n = |\mathbf{x}-\mathbf{x}_n|$ is the distance between the observer location and the $n$-th dipole.

The acoustic pressure is computed at the desired receiver locations and subsequently sound pressure levels (SPL), overall sound pressure levels (OASPL) , or other acoustic metrics are derived. The formulation assumes linear acoustic propagation in a homogeneous medium and neglects mutual hydrodynamic interaction between dipoles at acoustic frequencies.%, consistent with the free-field Green function approach adopted in Section \ref{subsec: Acoustic Formulation}.

% ---------------------------------------------------------------------- %
\subsubsection{Source reconstruction from OpenFAST}

The structural response obtained from OpenFAST provides nodal accelerations in the global inertial reference frame for the submerged structural nodes. Only wet nodes ($z<0$) are retained for acoustic radiation, as underwater sound generation originates from fluid--structure interaction below the free surface.
The acoustic source discretisation directly follows the structural finite-element output configuration defined in \texttt{SubDyn}. The adopted discretisation is:

\begin{itemize}
    \item \textbf{Monopile configuration:} 8 structural members with 5 output nodes per member, resulting in 25 submerged acoustic source nodes.
    
    \item \textbf{Floating configuration:} 9 structural members with 9 output nodes per member, resulting in 60 submerged acoustic source nodes distributed along columns and pontoons.
\end{itemize}

The structural mass associated with each node is inherited from the internal finite-element discretisation of \texttt{SubDyn} and subsequently used for equivalent force reconstruction.

Time-domain simulations are performed for 400 s. The initial 300 s are discarded to remove transient effects, leaving 100 s of statistically stationary data for spectral analysis. The simulations employ a time step of $\Delta t = 5\times10^{-4}$ s, corresponding to a sampling frequency of 2000 Hz and a Nyquist frequency of 1000 Hz.
The resulting frequency resolution is
$
\Delta f = \frac{1}{T} = 0.01 \ \text{Hz},
$
which ensures accurate resolution of both low-frequency structural dynamics and higher-frequency drivetrain harmonics.

For each submerged node and spatial direction, the acceleration signals are transformed into the frequency domain using a real-valued FFT formulation. Before transformation, the mean value was removed and a Hann window applied to reduce spectral leakage. The resulting complex acceleration spectra are subsequently converted into equivalent dipole forces through the effective mass formulation described in the following section.

% ---------------------------------------------------------------------- %
\subsubsection{Radiation-Impedance-Corrected Dipole Model}
\label{sec:rad}

The acoustic source reconstruction described above represents the submerged structure as a discrete distribution of pointwise dipoles. Although this approximation captures the global structural kinematics, it does not account for the finite cross-sectional dimensions of the structural members. In particular, offshore support structures are not sufficiently slender for their acoustic radiation to be represented solely by a line of compact sources at all frequencies.
Indeed, the finite size of the structural members introduces additional fluid--structure interaction effects that are absent from an ideal point-dipole formulation. These include radiation impedance, phase differences between the structural surface and the equivalent source location, and partial hydrodynamic shielding of high-frequency pressure fluctuations across the body surface. As a result, the acoustic field radiated by a three-dimensional structure exhibits a higher high-frequency decay than that predicted by a simple line distribution of dipoles.
To account for these effects within a reduced-order framework, the equivalent dipole forces are corrected using a frequency-dependent radiation impedance model.

\begin{equation}
\tilde{\mathbf{f}}_n(\omega)
=
\gamma(\omega,R_n)\;
m_{\mathrm{eff},n}\;
\tilde{\mathbf{a}}_n(\omega),
\label{eq:equivalent_force}
\end{equation}
where $\tilde{\mathbf{a}}_n(\omega)$ is the complex nodal acceleration obtained from OpenFAST, $m_{\mathrm{eff},n}$ is the effective structural mass associated with node $n$, and $\gamma(\omega,R_n)$ is a dimensionless frequency-dependent correction coefficient associated with the local cross-sectional radius $R_n$.

The effective structural mass is computed as

\begin{equation}
m_{\mathrm{eff},n}
=
\rho_{\mathrm{mat}}
A_n^*
\Delta z_n+\rho_{\mathrm{water}} A_n\Delta z,
\label{eq:structural_mass}
\end{equation}
where $A_n$ is the cross-sectional area enclosed by the outer boundary, $A_n^*$ is the net material area (i.e., cross-sectional area occupied by the solid structural material, excluding any hollow or water-filled region), and $\Delta z_n$ is the representative span length associated with the node.

The correction coefficient $\gamma(\omega,R_n)$ is derived from the analytical solution for the acoustic radiation of a harmonically oscillating rigid cylinder in an ideal fluid \cite{Junger1986}. Details of the derivation are provided in \ref{ap: gamma}. The correction coefficient is denoted as \(\gamma(\omega,R_n)\) to emphasise its dependence on frequency and local member radius, but for a homogeneous fluid, these quantities enter only through the local Helmholtz number \(kR_n=\omega R_n/c\), and consequently the coefficient can equivalently be written as \(\gamma(kR_n)\). The resulting radiation-impedance correction coefficient reads
\begin{equation}
    \gamma(kR_n)
    =
    \frac{4\mathrm{i}}{\pi (kR_n)^2 H_1^{(1)\prime}(kR_n)},
    \label{eq:gamma}
\end{equation}
where \(k=\omega/c\), \(R_n\) is the local external radius of the cylindrical member and \(H_1^{(2)\prime}\) denotes differentiation with respect to the argument. Hence, the correction depends on frequency and geometry only through the local Helmholtz number \(kR_n\).  %This formulation introduces the frequency dependence of the local radiation impedance into the discrete dipole representation while preserving the computational efficiency of the reduced-order model. 
From a physical perspective, the correction reproduces two key effects of finite-sized radiators:
\begin{itemize}
    \item \textbf{Low-frequency added-mass amplification}, where fluid inertia increases the effective radiated pressure relative to a free dipole source.
    \item \textbf{High-frequency attenuation and phase shielding}, arising from destructive interference and distributed surface radiation over the finite structural cross-section.
\end{itemize}
Consequently, the corrected formulation recovers the asymptotic decay behaviour expected for extended offshore structural members, which is not captured by an uncorrected line-source representation.
The correction is applied independently to each Cartesian acceleration component. For cylindrical members, $R_n$ corresponds to the external radius of the section, whereas an equivalent area-preserving radius is employed for rectangular pontoons.

It is important to note that this formulation does not attempt to reproduce the global hydrodynamic added-mass and radiation damping effects already included in \texttt{HydroDyn}. Instead, the objective is to incorporate the local frequency-dependent radiation physics governing acoustic emission into the equivalent source model.

Finally, the precision of the proposed formulation is assessed by comparison with a solution of a three-dimensional boundary element method. As will be detailed later in the validation section, Figure \ref{fig: validation} shows that the radiation-impedance correction improves the high-frequency behaviour of the model, reducing the overall acoustic error to below 1\% over the evaluated frequency range.

% ##################################################################### %
\subsection{Boundary Conditions: Method of Images}
\label{subsec: method of images}

\subsubsection{Image-Source Formulation}
The continuous and discrete dipole solutions, Eq.~\ref{eq: dipole solution} and Eq.~\ref{eq: linear superposition dipole solution}, respectively, are derived from the free-space Green function of the Helmholtz operator and are therefore strictly valid for an unbounded homogeneous medium. However, in the present problem, acoustic propagation takes place within a water layer bounded by a free surface at $z=0$ and a seabed located at $z=-H$.
To account for these horizontal boundaries, while preserving the analytical efficiency of the free-space formulation, the method of images is employed \cite{allen1979image}. The bounded-domain Green function is constructed as a superposition of free-space solutions associated with the physical dipoles and a set of virtual image sources generated by successive reflections at the boundaries. The total acoustic field is therefore written as
\begin{equation}
\tilde{p}(\mathbf{x},\omega)
=
\sum_{n=1}^{N_d}
\sum_{m=-\infty}^{\infty}
R_m\,
\tilde{p}_{\mathrm{free}}
(\mathbf{x};\mathbf{x}_{n,m},\omega),
\label{eq:image_method}
\end{equation}
where $\tilde{p}_{\mathrm{free}}$ denotes the free-space dipole solution, $\mathbf{x}_{n,m}$ is the position of the $m$-th image source associated with dipole $n$, and $R_m$ is the cumulative reflection coefficient corresponding to the reflection sequence.

The air--water interface is modelled using the classical pressure-release approximation \cite{urick1983principles}. Because of the strong acoustic impedance mismatch between air and water, the acoustic pressure is assumed to vanish at the free surface,
$
\tilde{p}=0 \, \text{at } z=0,
$
which is enforced through image sources with reflection coefficient $R_s=-1$, introducing the expected phase inversion upon reflection.
The seabed boundary at $z=-H$ is represented as a partially reflecting interface with constant reflection coefficient $R_b=0.5$, consistent with a moderately absorptive sandy sediment layer \cite{hamilton1980geoacoustic}. This simplified treatment introduces first-order bottom absorption effects while maintaining a computationally efficient propagation model suitable for comparative analysis between structural configurations.

% ---------------------------------------------------------------------- %
\subsubsection{Multiple Reflections and Infinite Image Series}

When there are two parallel boundaries, reflections occur successively between the surface and the seabed. Each newly generated image source produces additional reflections at the opposite boundary, resulting in an infinite sequence of image \cite{jensen2011computational, allen1979image} dipoles located at the following positions:
\begin{equation}
z_{n,m} = 2mH \pm z_n,
\end{equation}
where $m$ is an integer and $z_n$ is the vertical coordinate of the original dipole.
The corresponding reflection coefficient for each image is given by the product of surface and seabed reflection coefficients according to the number of reflections involved:
\begin{equation}
R_m = R_s^{\,n_s} R_b^{\,n_b},
\end{equation}
where $n_s$ and $n_b$ denote the number of reflections on the surface and the seabed, respectively.
The total acoustic field is therefore expressed as an infinite summation over all image sources. In practice, the series must be truncated after a finite number of reflections.

% ---------------------------------------------------------------------- %
\subsubsection{Truncation Error Convergence}

The infinite image series arising from successive reflections at the free surface and seabed must be truncated for practicality. Since each real dipole generates $(2N+1)$ sources when $N$ image reflections are included on each side, the total number of acoustic sources scales as $
N_{tot} = (2N+1)\,N_d,$
where $N_d$ denotes the number of real submerged dipoles. The computational cost therefore increases linearly with the number of retained image orders.

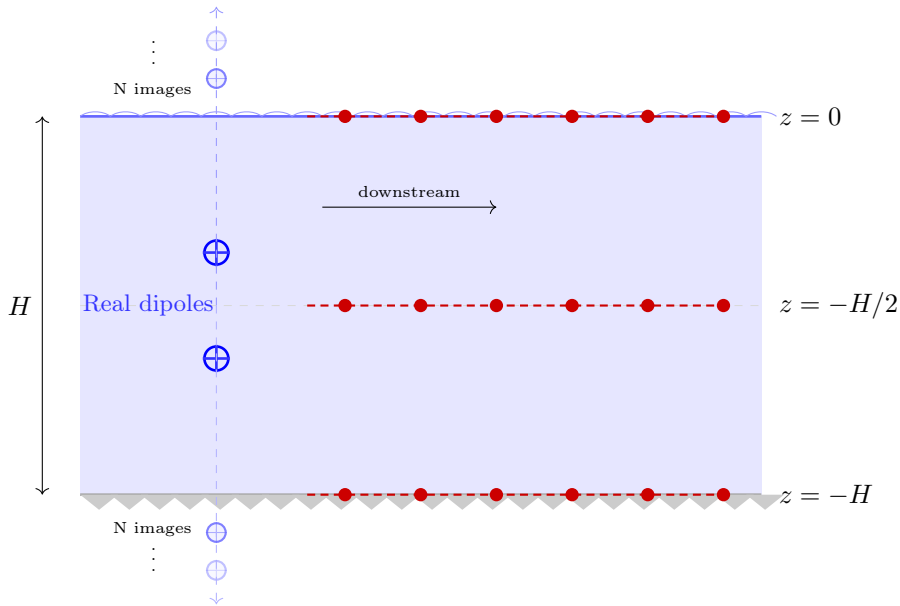
\begin{figure}[h]
\centering
\begin{tikzpicture}[
    scale=1.0,
    dipole/.style={circle, draw=#1, fill=#1!10, inner sep=0pt, minimum size=9pt},
    image/.style={circle, draw=#1, fill=#1!10, inner sep=0pt, minimum size=7pt, opacity=0.5},
    receiver/.style={circle, fill=#1, inner sep=0pt, minimum size=5pt},
]

%--- Dimensions
\def\Lx{9}       % horizontal extent
\def\H{5}        % water depth
\def\zs{1.8}     % source depth below free surface

%--- Ocean fill
\fill[blue!10] (0,0) rectangle (\Lx,-\H);

%--- Free surface
\draw[blue!60, line width=1pt] (0,0) -- (\Lx,0);
\foreach \i in {0,0.4,...,\Lx}
    \draw[blue!40] (\i,0) .. controls (\i+0.1,0.08) and (\i+0.3,0.08) .. (\i+0.4,0);
\node[right, font=\small] at (\Lx+0.1, 0) {$z=0$};

%--- Seabed
\draw[gray!70, line width=1pt] (0,-\H) -- (\Lx,-\H);
\foreach \i in {0,0.4,...,\Lx}
    \fill[gray!40] (\i,-\H) -- (\i+0.25,-\H-0.2) -- (\i+0.5,-\H) -- cycle;
\node[right, font=\small] at (\Lx+0.1, -\H) {$z=-H$};

%--- Mid-depth guide
\draw[gray!30, dashed, thin] (0,-\H/2) -- (\Lx,-\H/2);
\node[right, font=\small] at (\Lx+0.1, -\H/2) {$z=-H/2$};

%--- Depth annotation
\draw[<->, black, thin] (-0.5,0) -- (-0.5,-\H)
    node[midway, left, font=\small] {$H$};

%--- REAL DIPOLES
\pgfmathsetmacro{\xs}{1.8}
\node[dipole=blue, line width=1pt] (D) at (\xs,-\zs) {};
\draw[blue!80, line width=1pt] (\xs-0.18,-\zs) -- (\xs+0.18,-\zs);
\draw[blue!80, line width=1pt] (\xs,-\zs-0.18) -- (\xs,-\zs+0.18);
\node[dipole=blue, line width=1pt] (D) at (\xs,-\H+\zs) {};
\draw[blue!80, line width=1pt] (\xs-0.18,-\H+\zs) -- (\xs+0.18,-\H+\zs);
\draw[blue!80, line width=1pt] (\xs,-\H+\zs-0.18) -- (\xs,-\H+\zs+0.18);
\node[left, font=\footnotesize, blue!80] at (\xs+0.1,-\H/2) {Real dipoles};

%--- IMAGE SOURCES (above free surface, N=1 and N=2)
\foreach \imgz/\opa in {0.5/50, 1.0/25} {
    \node[image=blue, opacity=\opa/100, line width=0.8pt] at (\xs,\imgz) {};
    \draw[blue!60, opacity=\opa/100, thin]
        (\xs-0.14,\imgz) -- (\xs+0.14,\imgz)
        (\xs,\imgz-0.14) -- (\xs,\imgz+0.14);
}

%--- IMAGE SOURCES (below seabed, N=1 and N=2)
\foreach \imgz/\opa in {-5.5/50, -6.0/25} {
    \node[image=blue, opacity=\opa/100, line width=0.8pt] at (\xs,\imgz) {};
    \draw[blue!60, opacity=\opa/100, thin]
        (\xs-0.14,\imgz) -- (\xs+0.14,\imgz)
        (\xs,\imgz-0.14) -- (\xs,\imgz+0.14);
}

% %--- Reflection arrows
\draw[->, blue!40, dashed, thin] (\xs, -\H/2) -- (\xs, 1.45);
\draw[->, blue!30, dashed, thin] (\xs, -\H/2) -- (\xs, -\H-1.45);
\node[right, font=\tiny, black] at (\xs-1.5, +0.35) {N images};
\node[right, font=\tiny, black] at (\xs-1., +0.95) {$\vdots$};
\node[right, font=\tiny, black] at (\xs-1.5, -\H-0.45) {N images};
\node[right, font=\tiny, black] at (\xs-1., -\H-0.75) {$\vdots$};

%--- RECEIVER LINES
\def\xstart{3.0}
\def\xend{8.5}

% Free surface z=0
\draw[red!80!black, thick, densely dashed] (\xstart,0) -- (\xend,0);
\foreach \xr in {3.5,4.5,...,8.5}
    \node[receiver=red!80!black] at (\xr, 0) {};

% Mid-depth z=-H/2
\draw[red!80!black, thick, densely dashed] (\xstart,-\H/2) -- (\xend,-\H/2);
\foreach \xr in {3.5,4.5,...,8.5}
    \node[receiver=red!80!black] at (\xr, -\H/2) {};

% Seabed z=-H
\draw[red!80!black, thick, densely dashed] (\xstart,-\H) -- (\xend,-\H);
\foreach \xr in {3.5,4.5,...,8.5}
    \node[receiver=red!80!black] at (\xr, -\H) {};

%--- Downstream arrow
\draw[->, black, thin] (3.2, -1.2) -- (5.5, -1.2)
    node[midway, above, font=\tiny, black] {downstream};

\end{tikzpicture}
\caption{Geometry of the image method (xz plane). The real dipole (solid marker) is
placed at depth $z=-z_s$ and generates image sources above the free surface
and below the seabed through successive reflections. Receiver arrays at three
depths — free surface ($z=0$), mid-depth ($z=-H/2$), and seabed ($z=-H$) —
are distributed downstream for the convergence analysis.}
\label{fig: convergence analysis sketch}
\end{figure}

We include a convergence analysis to determine a suitable truncation level. Acoustic pressures are computed for an increasing number of images ranging from $N=1$ to $N=100$. The reference solution is defined using $N=100$ images, for which further increases in $N$ produced negligible variations.
The convergence assessment is conducted on multiple receiver lines located downstream of the rotor, spanning distances of 10 to 5000 m. Three horizontal receiver arrays were considered, as shown in figure \ref{fig: convergence analysis sketch}.
For each truncation level, the complex pressure field is evaluated for all receivers and for frequencies between 0 and 1000 Hz. The resulting pressure matrix is then compared to the reference solution $\mathbf{P}_{ref}$ using the relative Frobenius norm:
\begin{equation}
\varepsilon_N =
\frac{
\| \mathbf{P}_N - \mathbf{P}_{ref} \|_F
}{
\| \mathbf{P}_{ref} \|_F
}.
\end{equation}

\begin{figure}[H]
    \centering
    \includegraphics[trim=0 0 0 37, clip=True, width=0.75\textwidth]%scale = 0.65]
    {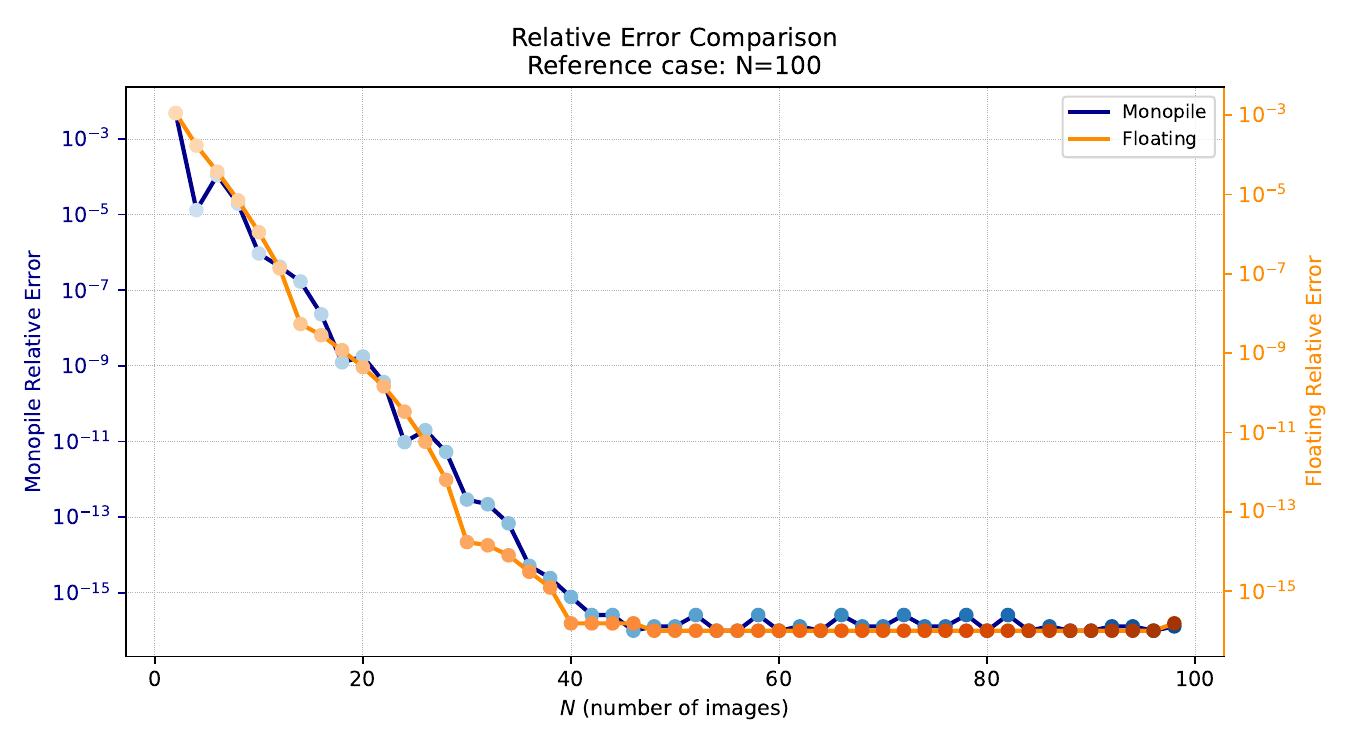}
    \caption{Convergence of the image source method. Relative Frobenius norm error of the pressure field as a function of the number of image reflections, using N = 100 as reference.}
    \label{fig: convergence analysis}
\end{figure}

Figure \ref{fig: convergence analysis} illustrates the decay of the relative error as a function of the number of image reflections. Exponential convergence is observed. For $N=30$, the relative error falls below $10^{-13}$, approaching machine precision (for double-precision arithmetic). Increasing the number of images beyond this value does not produce any significant improvement while increasing computational cost. Based on this analysis, $N=30$ image reflections are retained in all subsequent simulations, ensuring fully converged acoustic fields while maintaining computational efficiency.

Finally, let us note that the validity of the formulation of the present image method is restricted to horizontally planar air-sea interfaces and range-independent bathymetry.

% ##################################################################### %
\subsection{Validation}
\label{subsec: validation}

The proposed radiation-impedance-corrected dipole methodology was validated against a three-dimensional boundary element method \cite{kirkup2007boundary,ciskowski1991boundary,kirkup2019boundary} using a simplified monopile configuration in an unbounded fluid domain subjected to harmonic excitation. The acoustic pressure field was evaluated in the surrounding $xz$ plane and compared with the present reduced-order formulation.

\begin{figure}[h]
    \centering
    \includegraphics[trim = 0 20 0 35, clip = True, width=\linewidth]{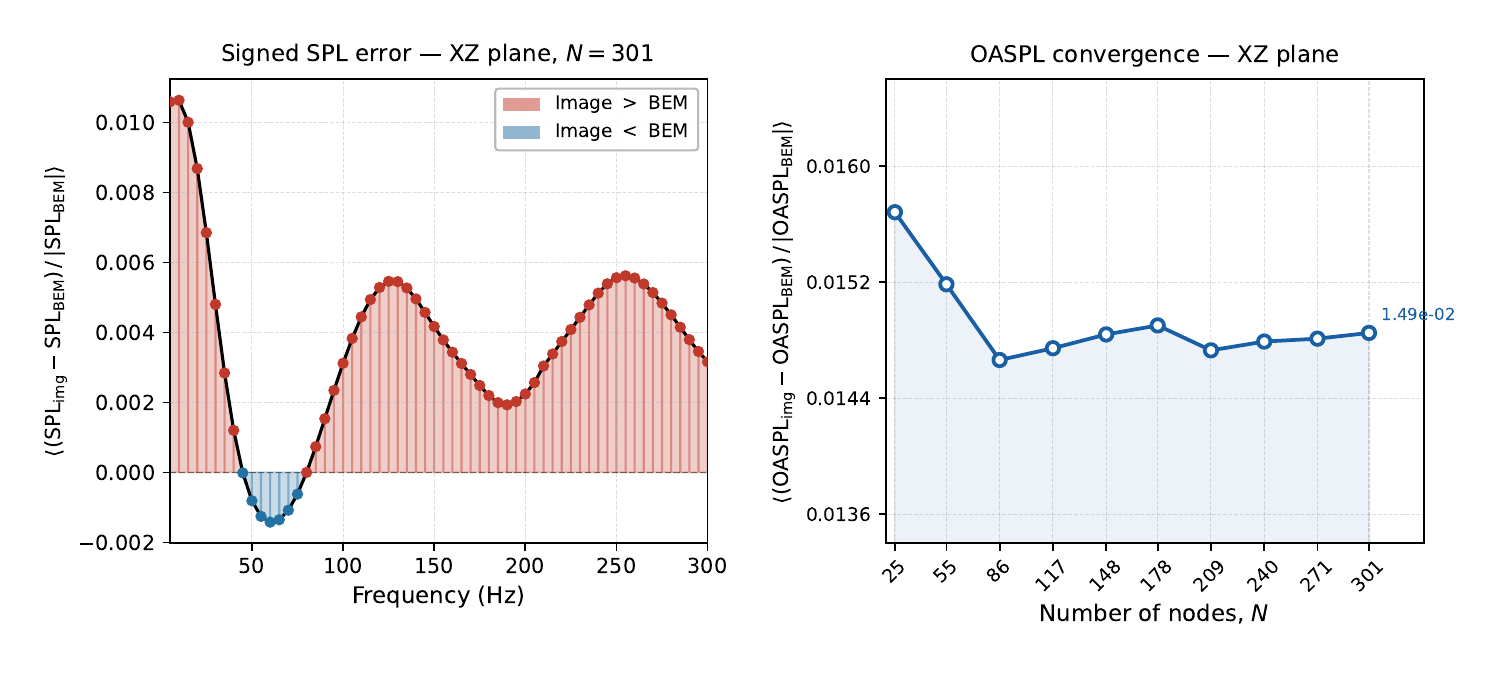}
    \caption{Validation of the proposed acoustic methodology against a three-dimensional Boundary Element Method (BEM) solution for a submerged monopile in free space. Left: mean signed relative SPL error over the evaluation plane as a function of frequency. Right: convergence of the maximum relative error with increasing number of equivalent dipole nodes.}
    \label{fig: validation}
\end{figure}

Figure~\ref{fig: validation} summarises the validation. The figure on the left shows the mean signed relative SPL error over the evaluation plane as a function of frequency. The results indicate that the radiation-impedance correction $\gamma(\omega,R)$ successfully captures the frequency-dependent decay of the radiated field and prevents the high-frequency over-prediction observed in uncorrected line-dipole formulations.
The figure on the right shows the convergence of the maximum relative error with an increasing number of equivalent dipole nodes used to discretise the monopile. Only minor variations are observed for the discretizations adopted in the present work, indicating mesh-independent acoustic predictions.
Overall, the comparison confirms that the proposed reduced-order methodology reproduces the dominant acoustic radiation behaviour while retaining low computational cost.

% Resultados
\section{Results}
\label{sec: results}

The acoustic response of monopile and floating offshore wind turbine is analysed for the following three cases:
\begin{itemize}
    \item \textbf{Monopile support structure} for water depth $H=30$ m.
    \item \textbf{The floating platform in shallow waters} for water depth $H=30$ m is included as a reference case to isolate the influence of structural typology on depth, and for ease comparison with the monopile case (the same depth is considered) .
    \item \textbf{Floating platform at deep depth of water} for depth of water $H=350$ m, representing the realistic depth of the floating configuration.
\end{itemize}
Unless otherwise stated, results are presented following the same ordering throughout the section.

% ##################################################################### %
\subsection{Structural Vibration and Radiation Mechanism}
\label{subsec: radiation mechanism}

\begin{figure}[h]
    \centering
    %left bottom right top
    \includegraphics[trim = 0 50 0 60, clip=true, width=\textwidth]{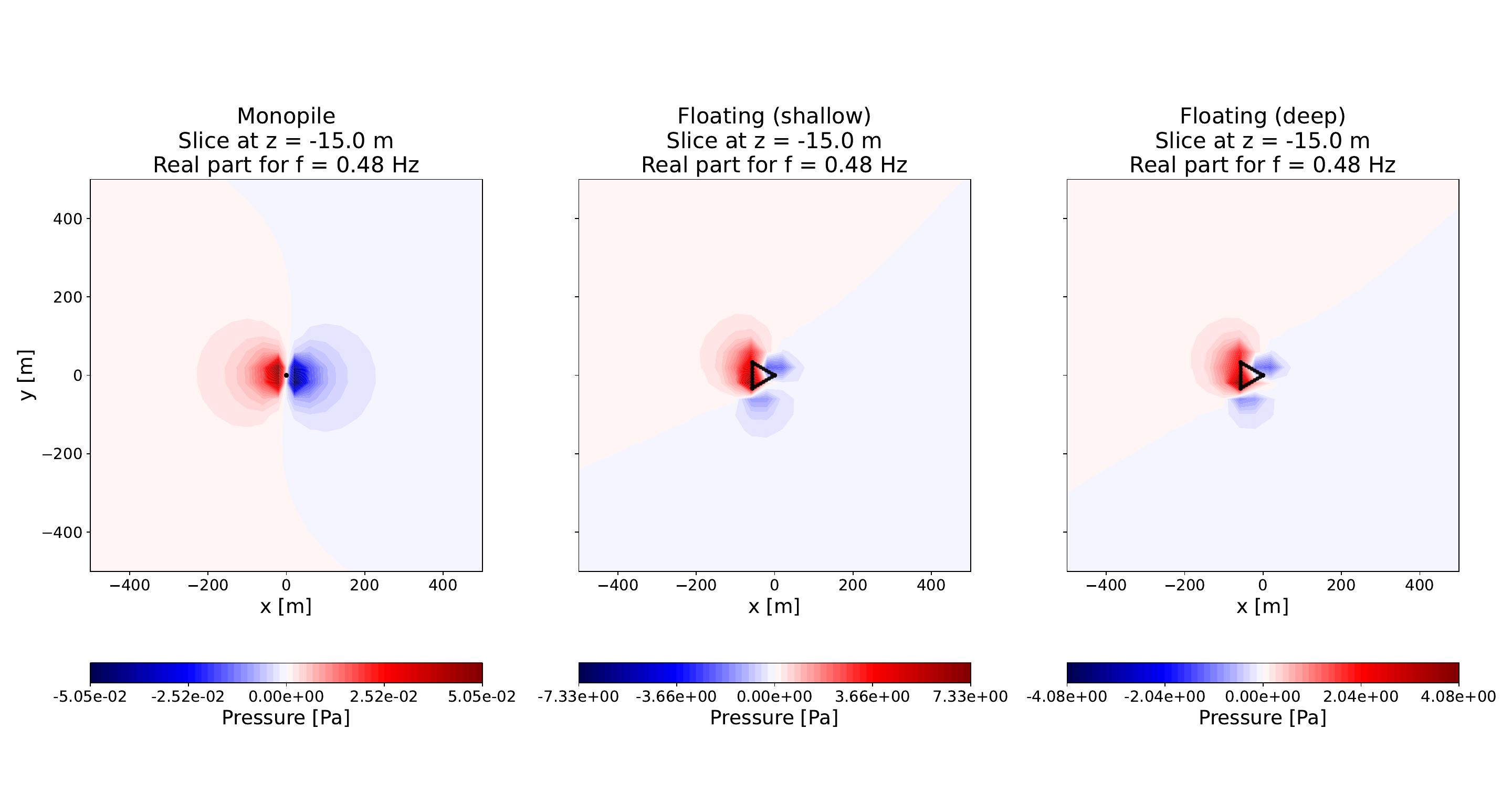}
    \caption{Real part of the acoustic pressure field in a horizontal plane at $z=-15$ m for the 3P blade-passing frequency (0.48Hz).}
    \label{fig: 3P real pressure}
\end{figure}

\begin{figure}[h]
    \centering
    \includegraphics[trim = 0 40 0 10, clip=true, width=\textwidth]{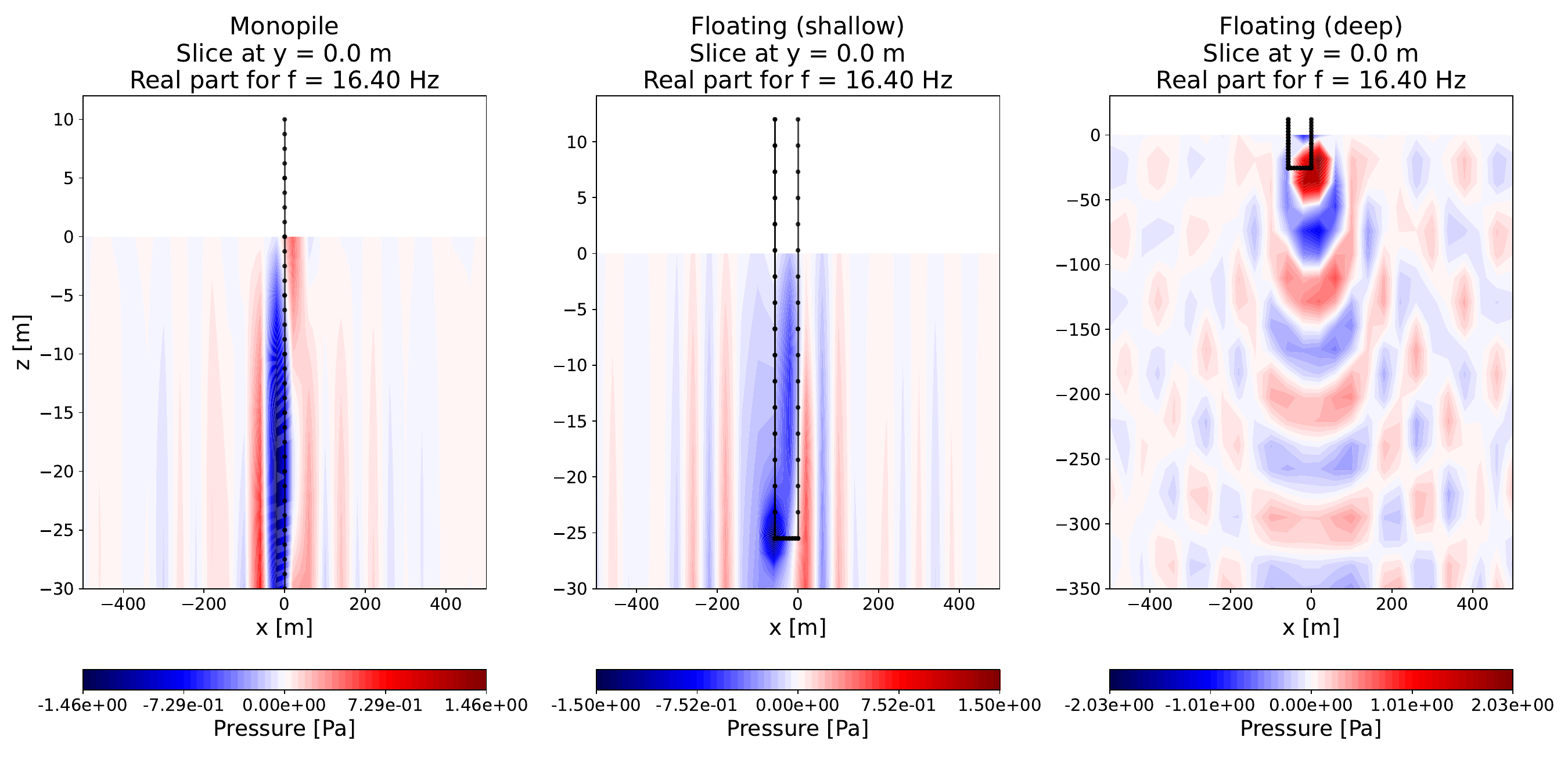}
    \caption{Real part of the acoustic pressure filed in the vertical streamwise plane ($y = 0$ m) for the first gear-mesh frequency GM1 (16.4 Hz).}
    \label{fig: GM1 real pressure}
\end{figure}

The radiated underwater sound field is governed by the coupling between structural vibration modes and the propagation characteristics of the surrounding fluid domain. At low frequencies, the acoustic response is primarily controlled by global structural motions, whereas at higher frequencies, both local structural excitation and environmental boundary effects become increasingly relevant. The pressure maps shown in Figures \ref{fig: 3P real pressure} and \ref{fig: GM1 real pressure} illustrate these two complementary regimes by comparing a low-frequency rotor-induced excitation with a higher-frequency drivetrain-related component

Figure \ref{fig: 3P real pressure} shows a horizontal slice at $z = -15$ m of the real part of the acoustic pressure field at 0.48 Hz, corresponding to the 3P blade-passing frequency. This tonal component excites the dominant fore-aft global bending mode of the turbine, producing a response that is essentially structural. For the monopile case, the radiated pressure field exhibits a clear dipolar pattern aligned with the streamwise direction, consistent with the oscillatory fore-aft motion of the tower-foundation system. This behaviour reflects the periodic aerodynamic thrust that acts on the rotor and the associated inertial loading transmitted through the support structure. The resulting antisymmetric directivity is characteristic of the force-driven dipole radiation, as expected from the theoretical formulation introduced in section 2.

In contrast, the floating platform configurations do not exhibit such a sharply defined dipolar structure. The pressure field becomes more spatially distributed and the main radiation axis is less pronounced. This difference indicates that the platform dynamics introduces additional coupled motions and load redistribution mechanisms that modify the effective source orientation and phase coherence along the submerged structure. Consequently, the floating concepts generate a more complex directivity pattern than the bottom-fixed monopile.

Figure \ref{fig: GM1 real pressure} shows the real part of the acoustic pressure field at 16.4 Hz, corresponding to the excitation frequency of the GM1 drivetrain identified in Figure \ref{fig: generator spectrum}, using a vertical slice at $y=0$ to highlight the interaction between the radiated wave and the boundaries. At this frequency, assuming a sound speed in water of approximately $c=1500$ m/s, the acoustic wavelength is close to 90 m. For the shallow-water cases, the wavelength is significantly larger than the water column height, constraining the propagation to a quasi-two-dimensional regime. As a result, acoustic energy is preferentially channelled in the horizontal direction, with wavefronts propagating nearly perpendicular to the structure axis. This behaviour is further reinforced by the elongated vertical geometry of the support structure, which acts as a line-type radiator in the water columns.

 For the floating case in deep-water, the propagation pattern changes substantially. Since the water depth is now several times larger than the acoustic wavelength, the field is no longer vertically constrained, and a fully 3D radiation pattern emerges, and clear constructive and destructive patterns arise from multiple reflections between the free surface and seabed, demonstrating the strong role of the environmental boundaries in shaping the radiated sound field once depth limitations are relaxed.

% ##################################################################### %
\subsection{Radiated Acoustic Spectra}
\label{subsec: spectra}

The frequency‐-dependent acoustic spectra provide insight into the physical mechanisms responsible for the generation and transmission of underwater noise. Although the spatial analyses presented in the following sections describe where the sound is radiated, the spectra identify which structural or drivetrain processes dominate the emitted energy. The narrowband sound pressure level is therefore used here as the primary descriptor of tonal and broadband content.

For each discrete frequency resolved by the Helmholtz formulation, the complex acoustic pressure $p(f)$ is converted into an equivalent root‑mean‑square value assuming harmonic steady‑state behaviour,
\begin{equation}
    p_\mathrm{rms}(f) = \frac{|p(f)|}{\sqrt2} \longrightarrow \text{SPL}(f) = 20 \log_{10}\!\left( \frac{p_\mathrm{rms}(f)}{p_\mathrm{ref}} \right),
    \label{eq: SPL}
\end{equation}
where $p_{ref}= 1\, \mu $Pa is the standard underwater acoustic reference pressure. This formulation is appropriate because each frequency component is solved independently in the frequency domain. Under the linear acoustic assumption, the total field can be interpreted as the superposition of uncoupled harmonic responses, so that each spectral line may be analysed separately before energetic recombination into broadband metrics, such as OASPL.

To account for propagation losses due to seawater absorption, we introduce an attenuation coefficient. Although geometric spreading is the dominant attenuation mechanism at the distances considered here, viscous and chemical relaxation processes progressively remove acoustic energy from the propagating wave, with stronger effects at higher frequencies. To represent this phenomenon, a frequency--dependent attenuation coefficient ‐ $\alpha (f)$ (in dB/m) is applied to the calculated SPL values as a linear loss proportional to the propagation distance.
\begin{equation}
    \text{SPL}_{\mathrm{corr}}(f,r) = \text{SPL}(f)-\alpha(f)r,
    \label{eq: absorption}
\end{equation}
where $\alpha(f)$ follows the classical Thorp absorption model \cite{Sehgal2010}. Although simplified, this correction captures the correct physical trend and prevents unrealistically persistent high‑frequency components from appearing in long‑range predictions.

Because the Helmholtz solution does not provide explicit ray trajectories, the exact travelled path of each reflected contribution cannot be uniquely determined. A conservative approximation is therefore adopted: the travelled distance is taken as the straight‑line distance between source and receiver, which represents a lower bound for the actual path length and yields a lower bound estimate of absorption losses. This ensures that attenuation is not over-estimated while still introducing physically significant dissipative effects into the spectral predictions.

\begin{figure}[h]
    \centering
    \includegraphics[trim = 0 0 0 25, clip=True, width=\textwidth]{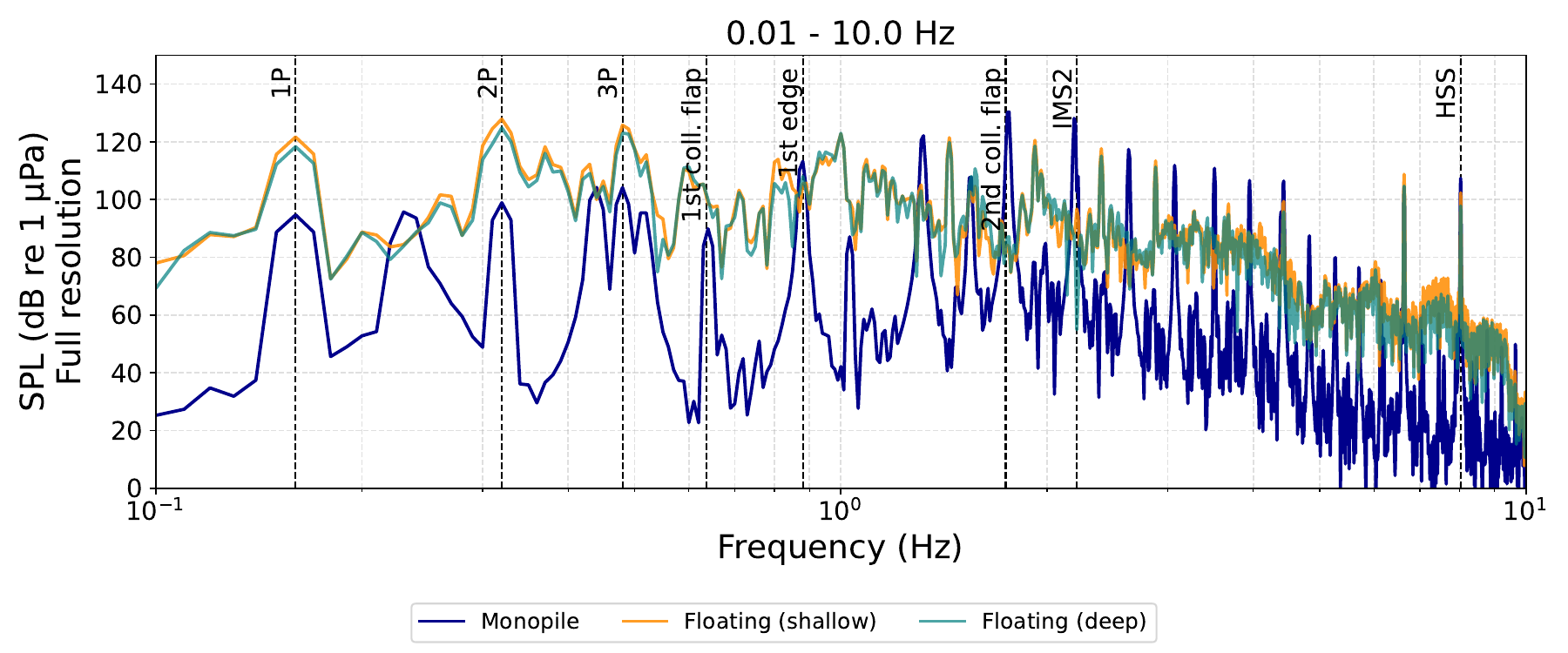}
    \caption{Narrowband Sound Pressure Level spectra in the low‑frequency range (0--10~Hz) evaluated at a receiver located 10~m from the turbine at $z = -15$~m. Dashed lines indicate blade passing harmonics (1P, 2P and 3P), characteristic drivetrain frequencies IMS and HSS (intermediate- and high-speed shaft) and structural modal frequencies (first collective flap, first edge and second collective flap modes) reported in \cite{bak2013dtu10mw, Borg2015LIFES50D12}.}
    \label{fig: low frequency spectra}
\end{figure}

Figure~\ref{fig: low frequency spectra} shows the narrowband SPL spectra for low frequencies (below 10 Hz) for the three configurations (monopile and floating in shallow and deep waters). The most significant feature is the substantially higher emission from both floating cases in the low‑frequency range ($f<1\, $Hz). This is a direct consequence of the additional rigid‑body degrees of freedom that characterise the dynamics of the floating platform and are absent in the monopile. The larger physical dimensions and operational mass of the semi‑submersible enhance the radiation efficiency at these very low frequencies, producing a continuous spectral hump that dominates the low‑frequency signature. In contrast, for frequencies above 1 Hz, the monopile turbine radiates noise more efficiently. In particular, the spectral peaks associated with the harmonics of the structural modes and the low-frequency components originating from the hub become more pronounced.

\begin{figure}[h]
    \centering
    \includegraphics[trim = 0 0 0 25, clip=True, width=\textwidth]{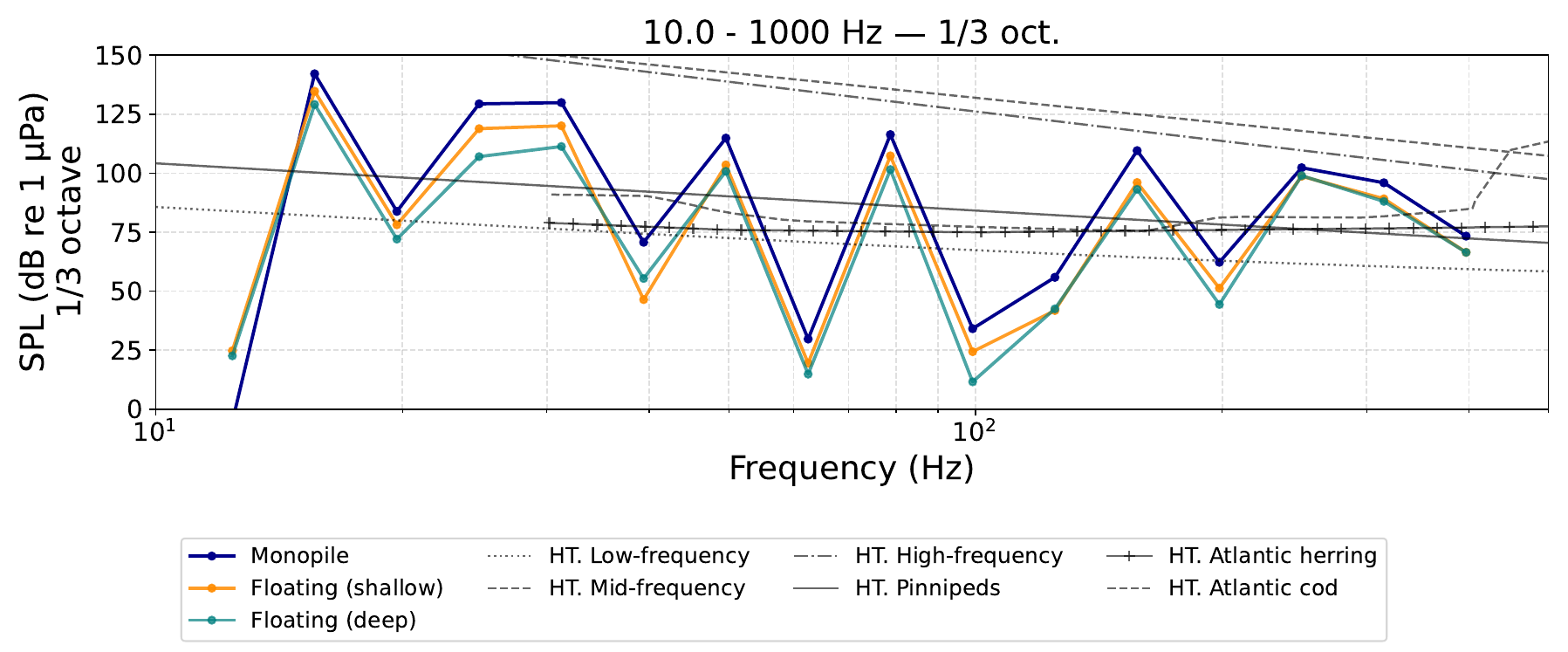}
    \caption{One‑third octave band Sound Pressure Level spectra (10--1000~Hz) evaluated at  a receiver located 10~m from the turbine at $z=-15$~m compared with marine fauna hearing thresholds from \cite{Tougaard2021, NMFS2024,Erbe2016}.}
    \label{fig: high frequency spectra}
\end{figure}

Figure~\ref{fig: high frequency spectra} presents the mid‑ to high‑frequency acoustic spectra (10 Hz to 500 Hz) for the three support configurations. In this case to improve clarity the SPL are expressed in the $1/3$ ‐ octave bands, according to standardised practices \cite{ISO266}. In contrast to the low‐frequency behaviour shown in Fig \ref{fig: low frequency spectra}, now the monopile configuration exhibits higher radiated levels in a substantial portion of the high‑frequency range. Although the $1/3$‑octave band representation inherently smooths narrowband tonal contributions, this trend becomes more evident in subsequent higher‑resolution analyses.

To provide context,  we include the hearing thresholds of marine species and show clear overlaps, suggesting that the turbine acoustic footprint can mask animal communications. Additional discussion is provided in Section~\ref{subsec: environmental impact}.

% ##################################################################### %
\subsection{Directivity Patterns}
\label{subsec: directivity}

\begin{figure}[H]
    \centering
    \includegraphics[trim = 0 0 0 30, clip=True, width=\textwidth]{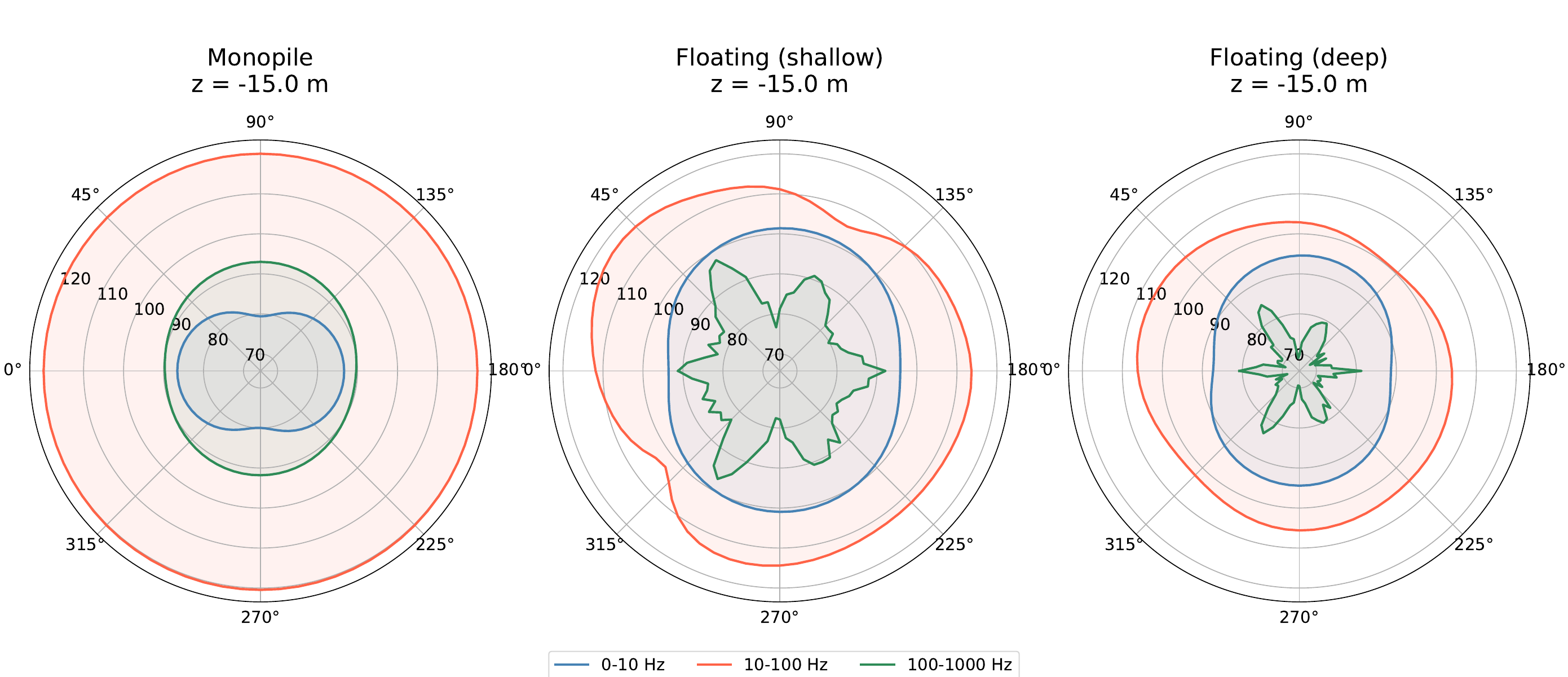}
    \caption{Polar distribution of OASPL at a radius of 500 m and depth z = -15 m, separated by frequency bands (0–10 Hz blue, 10–100 Hz orange, 100–1000 Hz green)}
    \label{fig: polar OASPL}
\end{figure}

To quantify the angular redistribution of the radiated sound field, Figure \ref{fig: polar OASPL} presents polar representations of the overall sound pressure level evaluated at a constant radius of $500$ m and at $z=-15$ m (mid-water depth for shallow water cases). For each angular position, the OASPL within a prescribed frequency interval is obtained by energetic summation of the corresponding narrowband sound pressure levels previously introduced in Section~\ref{subsec: spectra}, namely
\begin{equation}
    \text{OASPL}_{[f_1,f_2]}= 10 \log_{10}\left( \sum_{f=f_1}^{f_2} 10^{\text{SPL}(f)/10} \right).
    \label{eq: OASPL}
\end{equation}

This band-integrated metric allows the directivity associated with different physical excitation mechanisms to be analysed separately. In particular, the selected frequency ranges distinguish the low-frequency content dominated by global structural motions from the higher-frequency response primarily associated with drivetrain-related excitations.

For the monopile configuration, the low-frequency band preserves the dipolar radiation pattern already identified in Figure \ref{fig: 3P real pressure}. The alignment with the streamwise direction is consistent with the dominant fore–aft bending motion, confirming that the radiated field at low frequencies remains governed by a coherent force oscillation aligned with the direction of the incoming wind.

At higher frequencies, the monopile directivity becomes substantially more axisymmetric, with only a moderate preference for radiation in directions perpendicular to the streamwise axis. This behaviour is physically consistent with drivetrain-induced excitation, where the dominant forcing mechanisms originate from rotating shafts, bearings, and gear-mesh interactions within the nacelle \cite{Peeters2006,Kahraman1994}. Since these loads are related to rotational and torsional dynamics rather than a single translational bending mode, the resulting acoustic radiation is more spatially distributed and less directional.

A first comparison between the shallow- and deep-water floating cases reveals that the most immediate difference is the overall amplitude level, with the deep-water case exhibiting reductions of up to approximately 10 dB depending on the frequency band and direction. The larger propagation volume available in deep water promotes stronger three-dimensional geometric spreading, allowing acoustic energy to disperse more efficiently than in the shallow-water environment, where vertical confinement tends to retain energy within a reduced cross-sectional domain. Reduction in modal trapping of the water column in deep conditions may also contribute to lower received levels.

The most distinctive feature of floating configurations is not the amplitude reduction but the change in the angular radiation structure. In the low-frequency band, the principal emission axis is rotated relative to the streamwise direction, in agreement with the pressure maps shown in Figure \ref{fig: 3P real pressure}. This indicates that the dominant global motions of the floating platform are not purely fore–aft, but instead arise from coupled rigid-body and elastic responses that include significant rotational components, effectively reorienting the equivalent dipole source.

A more detailed comparison across frequency bands shows that the differences between configurations are not uniformly distributed. For the shallow floating case, deviations of the order of 10 dB with respect to the monopile are observed mainly in the intermediate frequency range (10 to 100 Hz), whereas for the deep-water configuration these differences extend to higher frequencies ($f > 100$ Hz). This behaviour is consistent with the spectral trends discussed in Figures \ref{fig: low frequency spectra} and \ref{fig: high frequency spectra}, where floating configurations exhibit enhanced low-frequency content, and with the stronger variations in directivity observed in that regime.

\begin{figure}[h]
    \centering\includegraphics[trim = 0 35 0 40, clip = true, width=\textwidth]{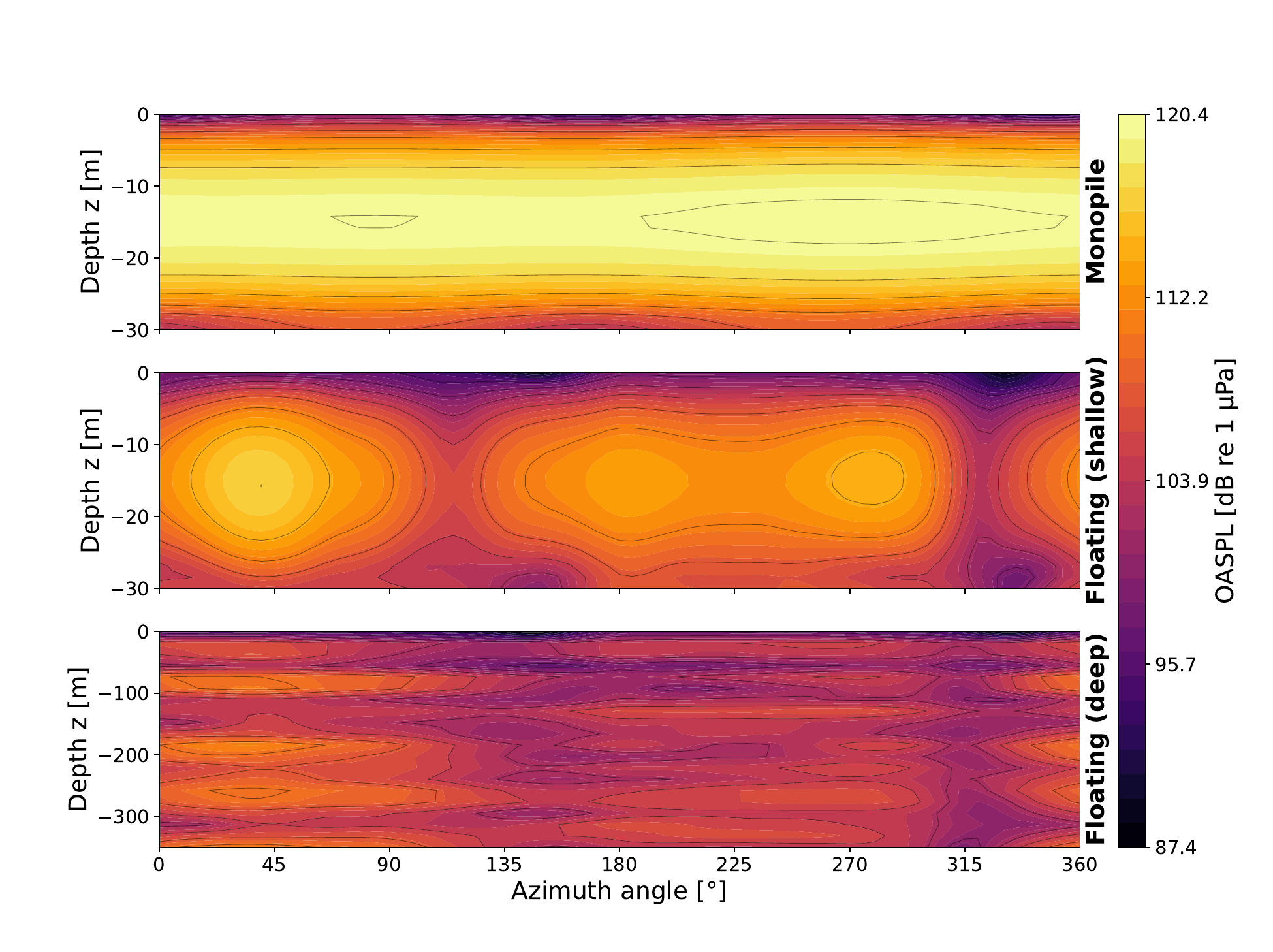}
    \caption{Cylindrical representation of OASPL as a function of azimuth and depth at a radius of 500 m.}
    \label{fig: cylinder OASPL}
\end{figure}

At higher frequencies, the floating cases exhibit a non-uniform directivity pattern characterised by four preferential radiation lobes located approximately at $\pm60^\circ$, $\pm120^\circ$, $0^\circ$ and $180^\circ$ from the upstream direction. This six-lobed structure suggests that the distributed geometry of the semi-submersible platform and the spatial arrangement of columns and pontoons introduce multiple coherent radiation centres. Their phase combination produces preferred constructive directions and suppressed sectors, resulting in a clearly anisotropic acoustic footprint. This non-uniform radiation is more clearly revealed in the cylindrical OASPL representation discussed in Figure \ref{fig: cylinder OASPL}.

Figure \ref{fig: cylinder OASPL} extends the polar directivity analysis of Figure \ref{fig: polar OASPL} by incorporating the vertical coordinate, thus providing a cylindrical representation of the OASPL as a function of azimuth and receiver depth at a constant radial distance. This additional dimension reveals how the angular radiation pattern evolves throughout the water column and clarifies the role of depth-dependent interference effects.

The monopile configuration shows a largely uniform radiation pattern with respect to depth, with only minor azimuthal variations. A slight enhancement is observed around $90^\circ$ and $270^\circ$, consistent with the preferential high-frequency emission directions associated with the orientation of the drivetrain. However, these variations remain weak, and the overall field can be considered nearly invariant along the vertical axis, apart from boundary-induced effects. This behaviour is consistent with the slender geometry of the monopile, which promotes relatively uniform radiation along its vertical extent.

In contrast, floating configurations exhibit a markedly more complex spatial structure. The preferred radiation directions become more pronounced and the azimuthal distribution is characterised by a more lobed pattern. In addition, a stronger dependence on depth is observed, reflecting the influence of the extended and distributed geometry of the floating platform.

\begin{figure}[h]
    \centering
    \includegraphics[trim = 0 50 0 0, clip = true, width=\textwidth]{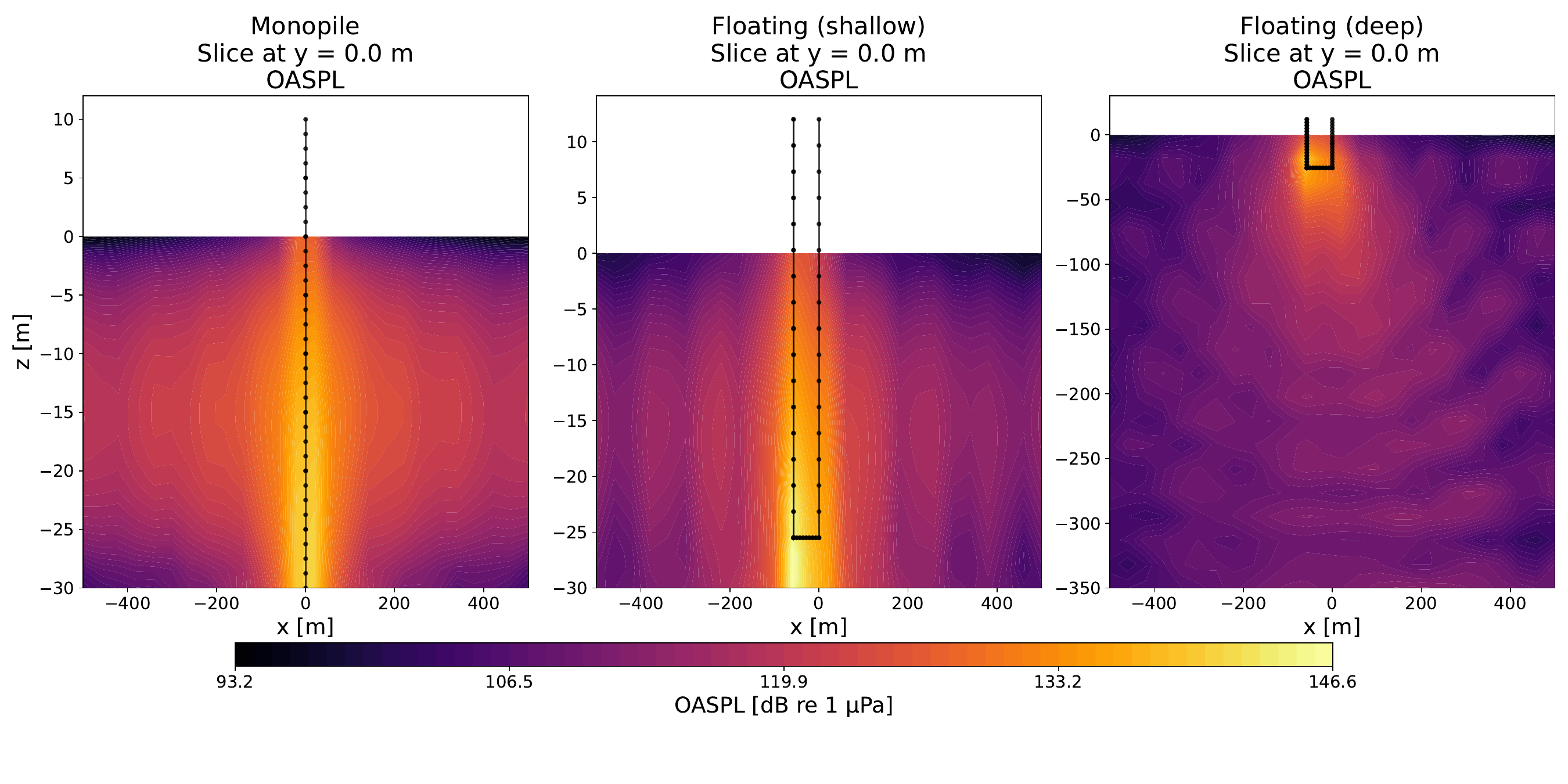}
    \caption{Vertical (x–z) distribution of OASPL in the streamwise plane $y = 0$~m.}
    \label{fig: OASPL xz}
\end{figure}

\begin{figure}[H]
    \centering
    \includegraphics[trim = 0 50 0 0, clip = true, width=\textwidth]{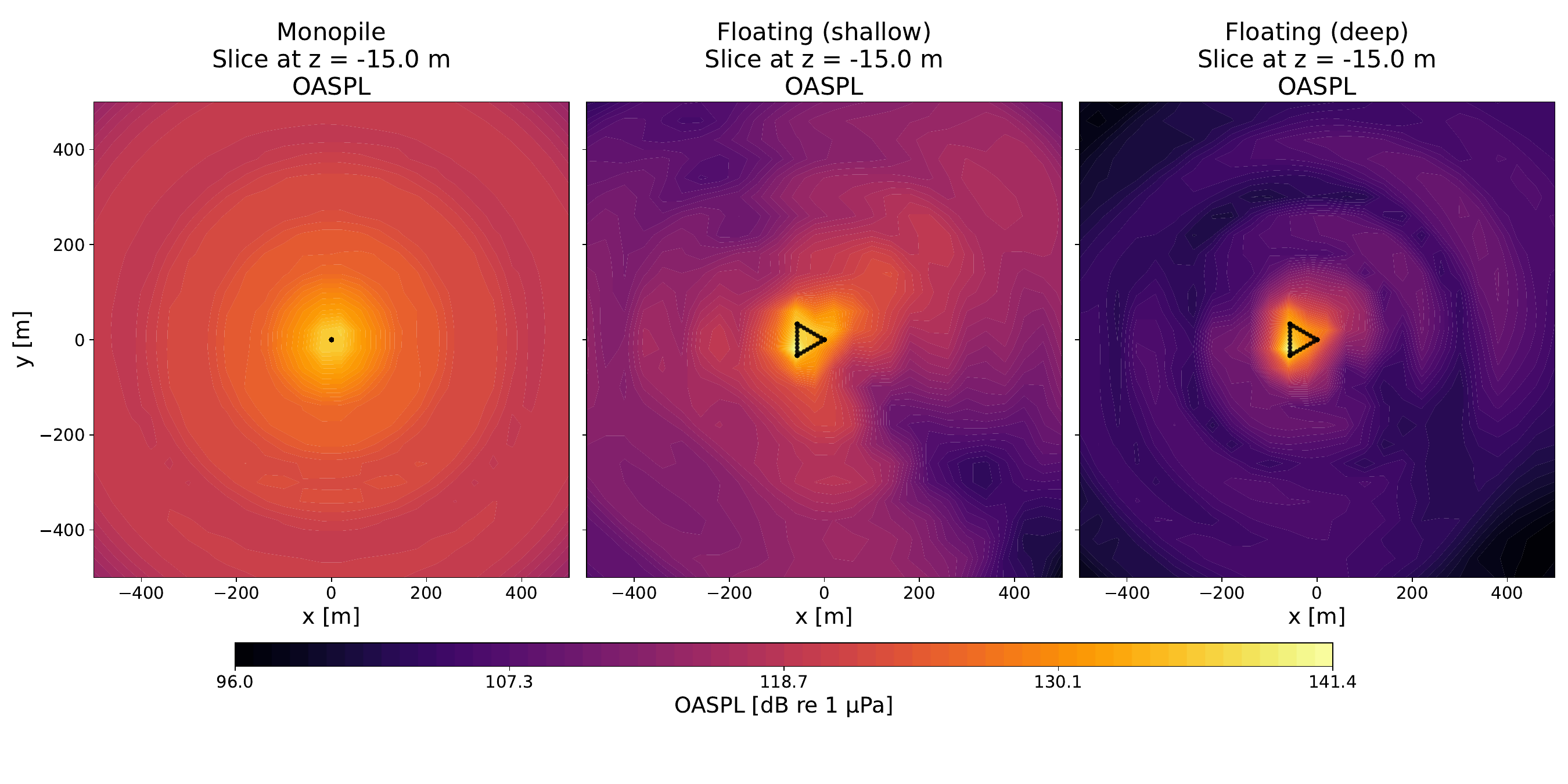}
    \caption{Horizontal (x–y) distribution of OASPL at $z = -15$~m.}
    \label{fig: OASPL xy}
\end{figure}

These observations are further clarified in Figures \ref{fig: OASPL xz} and \ref{fig: OASPL xy}, which provide planar views of the acoustic field and allow a more direct interpretation of the underlying propagation mechanisms.
In the vertical plane ($y=0$), the deep-water floating configuration clearly exhibits downward-directed radiation combined with pronounced interference fringes, confirming the role of multiple reflections between the seabed and the free surface. The alternating high- and low-level regions highlight the importance of coherent wave interactions, which persist even after broadband integration. In contrast, the shallow-water cases display a much smoother distribution, consistent with a propagation regime that is effectively constrained to two dimensions, with energy spreading predominantly perpendicular to the turbine axis.

In the horizontal plane ($z = -15$ m), the monopile configuration reveals a homogeneous radiation pattern, with generally higher overall sound pressure levels. However, floating configurations exhibit lower levels and a more irregular spatial distribution, with clear azimuthal modulation and localised interference effects. These patterns are particularly evident in the deep-water case, where the combined influence of structural complexity and boundary reflections leads to a more heterogeneous acoustic field.

Together, Figures \ref{fig: polar OASPL}-\ref{fig: OASPL xy} demonstrate that directivity in floating offshore wind turbines is inherently 3D. The acoustic footprint depends simultaneously on azimuth and depth, whereas the monopile response remains relatively simple and more vertically uniform. These results highlight that the receiver position in the water column may be as important as the horizontal distance when evaluating the exposure to underwater noise from floating wind systems.

\subsection{Distance Decay Analysis}
\label{subsec: distance decay}

\begin{figure}[h]
    \centering
    \includegraphics[trim= 0 50 0 0, clip=true, width=0.9\textwidth]
    {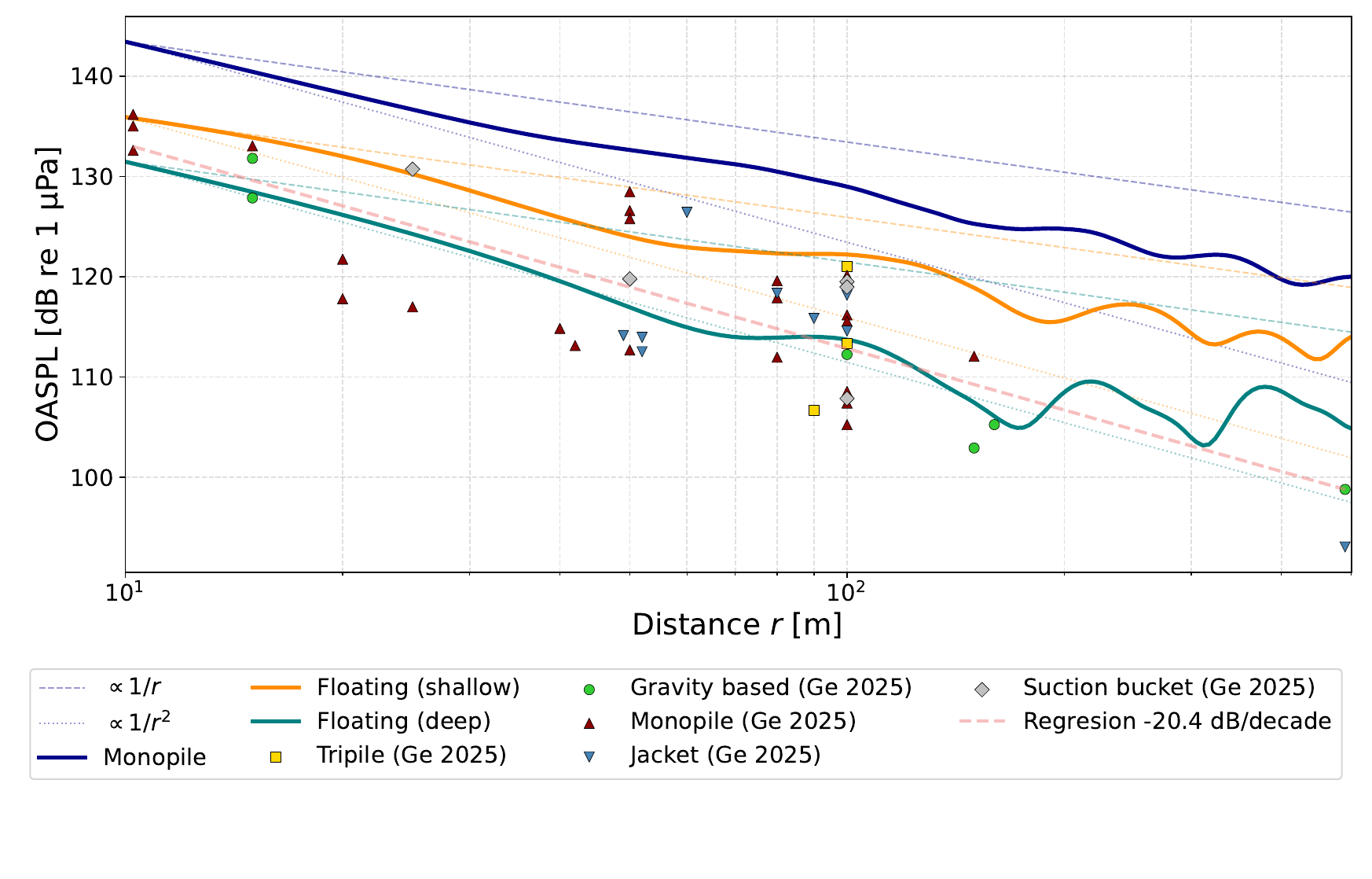}
    \caption{OASPL decay with distance along the downstream direction (y = 0, z = -15 m). Reference slopes corresponding to cylindrical ($1/r$) and spherical ($1/r^2$) spreading are shown. A regression fitted to the compiled dataset of offshore wind turbine measurements is also included, yielding a decay rate of -20.4 dB per decade.}
    \label{fig: Distance decay}
\end{figure}

Figure \ref{fig: Distance decay} examines the spatial attenuation of the radiated sound field as a function of the distance downstream from the turbine. The same definition of OASPL introduced in Eq. \ref{eq: OASPL} is used here, but now integrating the frequency range at various receiver locations that are distributed along the line defined by the intersection of the planes $y = 0$ and $z = -15$ m. This corresponds to the downstream receiver axis used in the propagation setup (equivalent to the mid-depth line in Figure \ref{fig: convergence analysis sketch} when $H = 15$ m). Receiver positions are logarithmically spaced between 10 m and 500 m in order to provide uniform visual resolution in the semi-logarithmic representation.

The first general observation is that the monopile configuration radiates higher overall levels than both floating configurations for all distances. Although the relative difference varies with distance, the bottom-fixed support consistently produces the strongest received OASPL, confirming its greater radiation efficiency already inferred from the directivity analysis.

For shallow-water cases, the attenuation rate lies between the classical limits associated with cylindrical and spherical spreading. For reference, ideal far-field geometric spreading predicts two limiting behaviours. In an unbounded 3D medium, the acoustic pressure amplitude decays as $|p|\sim 1/r$, which corresponds to a decay of $\text{OASPL} \sim 1/r^2$ and an OASPL reduction of 20 dB per decade. In contrast, an ideal 2D (cylindrical) propagation regime yields $|p|\sim 1/\sqrt{r}$, equivalent to $\text{OASPL} \sim 1/r$ and a level reduction of 10 dB per decade. The present results fall between these asymptotic behaviours, which is physically consistent with a partially confined propagation regime.

Two mechanisms explain this intermediate decay. First, shallow-water confinement modifies wave propagation whenever the acoustic wavelength becomes comparable to or larger than the water depth. In that regime, vertical propagation is restricted by the free surface and seabed, and the field progressively behaves as a reduced-dimensional or quasi-2D waveguide. A useful indicator is the ratio between depth and wavelength, or equivalently the non-dimensional parameter $kH$, where $k = 2\pi/\lambda$. When $kH\ll1$, only the lowest vertical mode can propagate efficiently and the field tends to horizontally guided spreading. When $kH\gg1$, several vertical modes are supported and the propagation approaches an unconstrained 3D regime. Thus, low frequencies in shallow water are the most prone to exhibit pseudo-2D behaviour.
Second, the geometry of the radiating structure also affects the decay with distance. In the monopile case, the submerged acoustic sources are arranged approximately along a vertical line. This configuration resembles, to first order, a finite line source normal to the horizontal observation plane. Since the ideal infinite line source is intrinsically a 2D radiator, the monopile naturally exhibits a stronger tendency toward $1/r$-type attenuation than the floating platform. The deviation from a perfect cylindrical law arises because the actual source is neither continuous nor infinite.

The shallow floating case also experiences depth confinement, but the decay departs more clearly from the monopile trend. The semi-submersible geometry includes horizontal pontoons and spatially separated columns, which generate radiation components not aligned with the vertically guided propagation path. As already observed in Figure \ref{fig: GM1 real pressure}, some structural elements radiate energy directly toward the seabed and into oblique directions, enhancing 3D spreading close to the platform. Consequently, the near-source region displays a stronger spherical-like tendency, while at larger distances the shallow-water confinement becomes increasingly dominant and the attenuation gradually approaches a pseudo-2D regime.

The deep-water floating configuration exhibits a more complex decay behaviour compared with the simpler spreading laws discussed previously. In the near-source region, the attenuation closely follows the classical spherical spreading law, with an effective decay proportional to approximately $1/r^2$, consistent with a predominantly three-dimensional propagation. However, as distance increases, the decay gradually departs from this behaviour and tends towards an approximate dependence $1/r$. This transition indicates the onset of confinement effects within the propagation domain, where repeated interactions with the free surface and seabed progressively modify the free-field propagation characteristics and promote a more waveguide-like acoustic behaviour that sustains acoustic energy over longer ranges.

To provide context, in Figure \ref{fig: Distance decay} the results are compared with field measurements \cite{Ge2025} for a range of offshore wind turbine configurations. The predicted levels for floating configurations fall within the acoustic measurements. Our  monopile results exhibit higher sound pressure levels than the monopile cases included in the measurements. Nevertheless, most of the turbines considered in the measures \cite{Ge2025} are rated below 5 MW, while the present study considers a turbine of 10 MW. The closest agreement is observed with the 8 MW turbine reported in the measurements, for which the predicted levels are only 5 dB higher, a difference that is consistent with the expected increase in radiated acoustic power associated with larger turbine size and structural loading.

In general, the comparison confirms that the predicted decay behaviour lies within the range of previously observed offshore wind turbine acoustic propagation measurements, while also highlighting the combined influence of source size, environmental confinement, and frequency-dependent propagation mechanisms.

% ##################################################################### %
\subsection{Impact of Atmospheric Turbulence}
\label{sec:turbulent_inflow}

The results presented so far were obtained under steady uniform inflow, which allowed cleaner spectra and facilitated the analysis. In this section, we include  atmospheric turbulence and repeat the simulations to assess its impact on underwater noise. We used a turbulent wind field generated with \texttt{TurbSim}~\cite{TurbSim_manual}, and a turbulent inflow that follows the IEC Kaikal spectral model (class B, turbulence intensity $\mathrm{TI}\approx15\%$ at hub height) with a mean wind speed of $11.4~\mathrm{m/s}$, matching the rated condition of the DTU 10 MW turbine. All other simulation parameters remain unchanged.

\begin{figure}[h]
\centering
\includegraphics[trim = 0 0 0 25, clip=True, width=\textwidth]{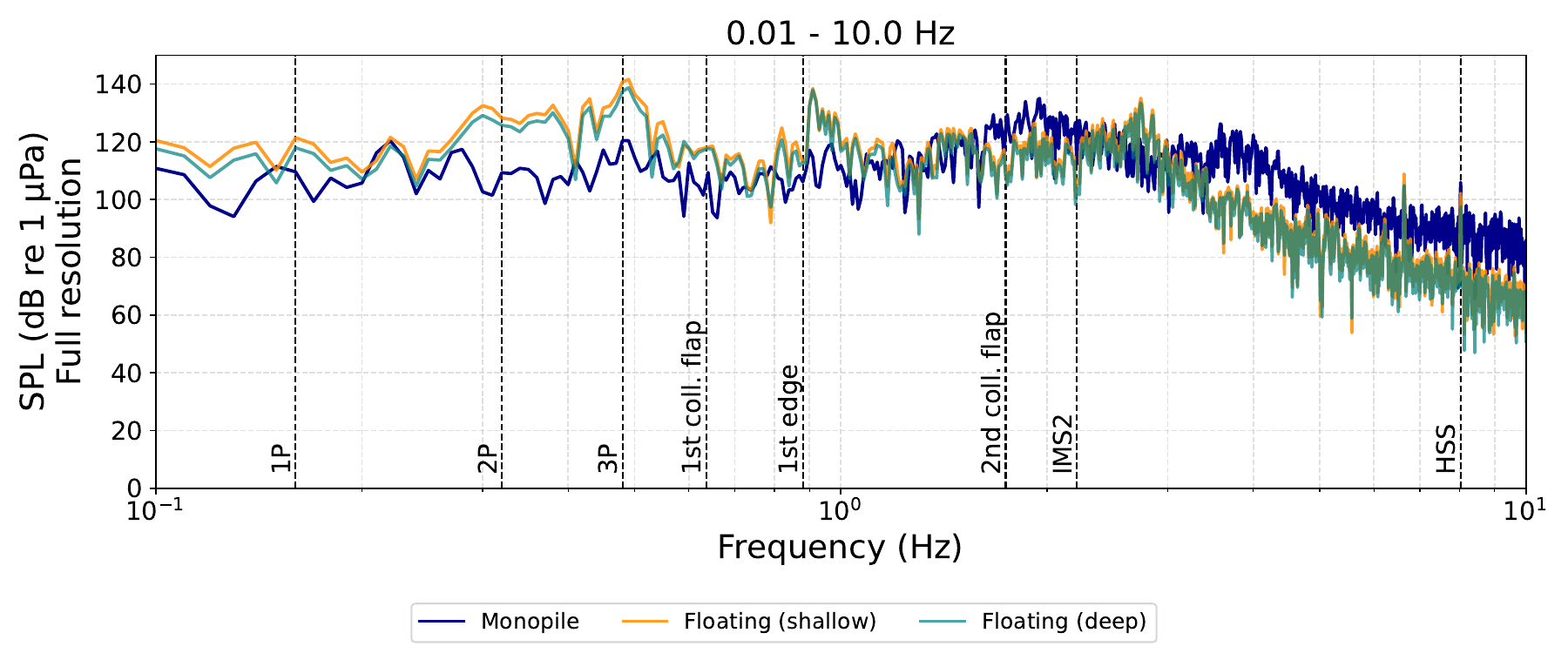}
\caption{Narrowband Sound Pressure Level spectra in the low‑frequency range (0--10~Hz) evaluated at a receiver located 10~m from the turbine at $z = -15$~m for a turbulent inflow. Dashed lines indicate blade passing harmonics (1P, 2P and 3P), characteristic drivetrain frequencies IMS and HSS (intermediate- and high-speed shaft) and structural modal frequencies (first collective flap, first edge and second collective flap modes) reported in \cite{bak2013dtu10mw, Borg2015LIFES50D12}.}
\label{fig:turbulent_spectrum}
\end{figure}

Figure~\ref{fig:turbulent_spectrum} compares the low-frequency narrowband SPL spectra (0–10~Hz) for the monopile configuration under steady and turbulent inflow at the same receiver location as in Figure~\ref{fig: low frequency spectra}. The most prominent effect of turbulence is the broadening of the originally sharp tones (blade-passing harmonics and low-frequency structural modes), which transform into wider humps. In addition, the overall sound pressure level increases, with local increases of up to $10~\mathrm{dB}$ in some peaks. The rise is most pronounced at low frequencies, while at higher frequencies the effect is less pronounced. Despite these changes, the previous conclusions remain unchanged: enhanced low-frequency radiation of floating platforms, higher monopile noise efficiency at higher frequencies, and the role of water depth in shaping noise porpagation underwater.

% Discusión
\section{Discussion}
\label{sec: discussion}

\subsection{Global Physical Interpretation}
\label{subsec: physical interpretation}

The results presented in Section 3 consistently show that the underwater acoustic signature of offshore wind turbines is strongly governed by the interplay between structural dynamics and the environment. Across all configurations, the dominant radiation mechanism remains dipolar, driven by inertial forces transmitted from the vibrating structure to the surrounding water. However, the efficiency and spatial characteristics of this radiation vary significantly with the typology of the support structure.

At low frequencies, floating configurations exhibit substantially higher radiated levels than the monopile case. This behaviour is directly associated with the presence of additional rigid-body degrees of freedom and the significantly larger structural mass of the floating platform. These factors enhance global motions and increase the effective force transmitted to the fluid, resulting in a stronger low-frequency acoustic footprint. In contrast, the monopile, constrained by seabed fixation, suppresses rigid-body motion and, therefore, radiates less efficiently in this frequency range.

At higher frequencies, the trend reverses, with the monopile configuration exhibiting comparatively higher sound pressure levels. In this regime, the structural response is dominated by local vibrations and drivetrain-induced excitations, which are more efficiently transmitted through the monopile structure. The floating platform, by contrast, distributes these excitations over a larger structure, leading to partial energy redistribution and reduced radiation efficiency at higher frequencies.

Beyond structural vibrations, the results demonstrate that the depth of the water plays a critical role in shaping the acoustic field. In shallow-water conditions, the acoustic wavelength is often larger than the height of the water column, leading to vertically constrained propagation and quasi-2D spreading. In deep-water environments, this constraint is removed, allowing fully 3D propagation and promoting stronger geometric spreading. As a result, both the amplitude and the spatial distribution of the acoustic field are strongly modulated by depth, independently of the source characteristics.

\subsection{Environmental Impact}
\label{subsec: environmental impact}

%The spectral results presented in Section~\ref{subsec: spectra} show that, for both support configurations, the radiated acoustic energy overlaps with the hearing thresholds of several marine species, particularly within the mid‑ to high‑frequency bands illustrated in Fig.~\ref{fig: high frequency spectra}. This confirms that, at least under near‑field conditions, the emitted sound is in principle detectable and therefore potentially relevant from an environmental perspective.
The spectral results presented in Section~\ref{subsec: spectra} show that, for both support configurations, the radiated acoustic energy overlaps with the hearing thresholds of several marine species, particularly within the mid- to high-frequency bands illustrated in Fig.~\ref{fig: high frequency spectra}. This confirms that, at least under near-field conditions, the emitted sound is in principle detectable and therefore potentially relevant from an environmental perspective. Detectability implies that turbine noise may be perceived by marine fauna, which can promote avoidance responses in some cases, but may also lead to acoustic masking when radiated levels overlap with frequency bands used for communication, orientation, prey detection, or predator avoidance. 

A clear distinction emerges between the two support structure typologies. The monopile configuration radiates higher sound pressure levels in the mid‑ and high‑frequency range (Fig.~\ref{fig: high frequency spectra}), which coincides with the most sensitive hearing regions of many marine vertebrates. Pinnipeds (seals) and odontocetes (dolphins, porpoises) possess auditory bandwidths that typically extend down to $40$--$150$~Hz \cite{Erbe2016, Southall2007, NMFS2024}, which means that a large part of the monopile ‐-dominated spectrum lies within their functional hearing range. This suggests a greater potential for short‑range detection and, depending on the duration and level of exposure, a higher probability of behavioural or physiological effects \cite{Southall2019}. In contrast, the floating configuration concentrates a significantly larger fraction of its acoustic energy at low frequencies (Fig.~\ref{fig: low frequency spectra}), where two mitigating factors substantially reduce ecological concern.

First, the very low‑frequency content of floating turbines ($<10$~Hz) falls partially or entirely outside the effective hearing range of the majority marine species. For odontocetes and pinnipeds, the acoustic energy around $5$~Hz lies several octaves below their lower sensitivity limit and is not expected to elicit an auditory response \cite{Kastelein2009}. Even low‑frequency specialist cetaceans, whose estimated hearing range reaches approximately $7$~Hz \cite{Southall2007} and that can use bone ‐ signalling pathways \cite{Cranford2015}, operate at the extreme lower edge of their sensitivity. Benthic invertebrates and crustaceans respond primarily to particle motion rather than acoustic pressure, but their mechanoreceptive systems are sensitive to substrate‑borne vibrations in the $5$ -- $200$~Hz band \cite{Roberts2015,Roberts2016,Roberts2017}. However, such particle‑motion responses are highly localised and decay rapidly with distance from the source, so they are relevant only in the immediate vicinity of the foundation.

Second, the infrasonic frequency band ($<5$~Hz) is typically dominated by strong natural background noise. Persistent ambient sources such as microbaroms generated by non-linear wave–wave interactions (peaking around $0.2$~Hz) \cite{Longuet-Higgins1950}, distant shipping, and seismic activity produce broadband noise floors that can exceed $120$--$140$~dB re $1\,\mu$Pa in the $1$--$10$~Hz band \cite{urick1983principles}. Wind‑driven surface dynamics further increases ambient levels at these frequencies \cite{Andrew2002, Wenz1962}. Consequently, the turbine‑induced signal in the lowest frequency range is often masked by natural ambient noise, reducing its detectability even for the few organisms nominally sensitive to such frequencies.

It should be noted that the present analysis does not include ambient background noise. Background noise is strongly site-dependent and varies with weather conditions, season, time of day, vessel activity, biological sources, and sea state. For this reason, it has not been included here, allowing the turbine contribution to be assessed in isolation. If a specific deployment site, period of the year, and time of day are considered, representative background-noise spectra can be added to the frequency-domain results presented here to evaluate the effective signal-to-noise ratio and the resulting audibility or masking potential under realistic environmental conditions.
%
%It should be noted that the effective acoustic footprint may extend beyond the frequency bands explicitly resolved by the present model. 
Additionally, in this work, we did not include Doppler effects associated with structural motion, time‑varying boundary conditions induced by surface waves, or more realistic representations of the seabed that can redistribute acoustic energy between frequencies and modify propagation characteristics. These effects could potentially shift part of the radiated energy into more sensitive auditory bands or alter its spatial extent, and these effects will be studied in future work.

From a design perspective, the findings presented here reinforce the idea that the choice of support structure has a direct influence on underwater noise emissions. Consequently, both the structural configuration and the site ‐specific environmental conditions (water depth, ambient noise climate, presence of sensitive species) should be considered jointly when evaluating the potential acoustic impact of offshore wind developments.

\section{Conclusions}
\label{sec: conclusions}

This work presented a physics-based vibroacoustic framework to predict operational underwater noise from offshore wind turbines and applied it to a comparative analysis of monopile-supported and floating 10 MW configurations. The approach combines aero-hydro-servo-elastic time-domain simulations with a frequency-domain acoustic formulation based on equivalent dipole sources and Green's function solutions, including the effect of water-column confinement through image-source representations.  The following conclusions are drawn:

%The results show that the support configuration has a significant influence on both the level and spatial distribution of the radiated underwater sound. Floating turbines generate stronger low-frequency emissions, mainly due to the contribution of platform rigid-body motions, whereas monopile-supported turbines tend to radiate more efficiently at higher frequencies associated with drivetrain-related excitations. Differences in directivity and three-dimensional radiation patterns were also observed, with floating configurations producing more complex acoustic fields. In addition, water depth was shown to play an important role in controlling propagation regimes and overall sound levels, particularly for floating platforms in shallow-water environments.

%Overall, the proposed methodology provides a useful predictive tool for assessing operational underwater noise during the design stage of offshore wind projects. By linking turbine dynamics, support-structure response, and acoustic radiation, the framework can support the comparison of alternative designs, the identification of dominant noise-generation mechanisms, and the development of environmentally informed mitigation, monitoring, and regulatory strategies for future offshore wind deployment.

\begin{itemize}
%    \item A physics-based vibroacoustic framework has been developed to predict structure-borne underwater noise from offshore wind turbines, enabling a consistent comparison between monopile and floating support configurations under identical operating conditions.
%
    \item The results demonstrate a clear frequency-dependent behaviour: floating platforms radiate significantly higher levels at low frequencies due to additional rigid-body degrees of freedom and larger structural mass, whereas monopile configurations exhibit higher radiation efficiency at mid- to high frequency ranges.

\item The directivity patterns reveal a clear dependence of the radiated sound field on the support configuration. The monopile case exhibits a smooth and nearly axisymmetric response, whereas the floating configurations produce more directional and irregular radiation patterns, especially in the 100--1000 Hz band. This indicates that the additional rigid-body motions and coupled platform dynamics of floating turbines introduce more complex three-dimensional acoustic radiation features, which may lead to direction-dependent received levels even at the same radial distance.

\item The depth of the water is identified as a key parameter controlling the acoustic propagation. Shallow-water conditions promote vertically constrained, quasi-two-dimensional propagation, while deep-water environments enable fully three-dimensional spreading and reduce the overall received levels. This effect is particularly relevant for floating configurations, for which shallow-water conditions lead to variations of up to 7\% ($\sim$9 dB) in OASPL with respect to the corresponding deep-water cases.

    \item Spectral analysis reveals an overlap between turbine-generated noise and the hearing ranges of several marine species in both configurations. This overlap is more pronounced for monopile systems at mid- and high frequencies, whereas floating platforms concentrate energy in lower-frequency bands that may be partially outside the biological sensitivity ranges.

    \item The predicted acoustic response is subject to modelling assumptions, particularly regarding drivetrain excitation or environmental conditions (e.g., site specific bathymetry). %Additional physical mechanisms not included in the present study can potentially redistribute acoustic energy and should be considered in future work.

    \item The proposed methodology provides a fast predictive tool for early-stage design and environmental assessment, supporting the evaluation of alternative support structures and site conditions with explicit consideration of their acoustic footprint.
    
\end{itemize}

\section*{Acknowledgments}
This research has received funding from the European Union (ERC, Off-coustics, project number 101086075). However, the views and opinions expressed are those of the authors alone and do not necessarily reflect those of the European Union or the European Research Council. Neither the European Union nor the granting authority can be held responsible for them.

All authors gratefully acknowledge the Universidad Politécnica de Madrid (www.upm.es) for providing computing resources on Magerit Supercomputer and the computer resources at MareNostrum and the technical support provided by Barcelona Supercomputing Center (projects RES-IM-2024-1-0003 and RES-IM-2025-1-0011).

% Bibliografía
\bibliographystyle{elsarticle-num}
\bibliography{refs.bib}

%\section*{Appendices}
\appendix

% ##################################################################### %
\section{Continuous Acoustic Formulation}
\label{ap: Acoustic Formulation}

The acoustic model employed in this work is based on the linearised compressible Euler equations with external forcing \cite{wagner1996wind}. Structural vibrations are represented as localised body forces acting on the surrounding fluid such that

\begin{equation}
F_i(\mathbf{x},\mathbf{y},t)=f_i(t)\delta(\mathbf{x}-\mathbf{y}),
\end{equation}
where $\mathbf{y}$ denotes the source position and $\delta$ is the Dirac delta distribution.

Assuming small perturbations about a quiescent mean state and isentropic compressibility, the governing equations reduce to the classical inhomogeneous acoustic wave equation for pressure fluctuations \cite{pierce2019acoustics}

\begin{equation}
\frac{1}{c_0^2}\frac{\partial^2 p'}{\partial t^2}
-\nabla^2 p'
=
-\nabla\cdot\mathbf{F},
\label{eq: wave equation}
\end{equation}
where $c_0$ is the speed of sound in water.
We apply the temporal Fourier transform using the convention
\[
p'(\mathbf{x},t)=\tilde{p}'(\mathbf{x},\omega)e^{-i\omega t},
\]
which yields the frequency-domain Helmholtz equation
\cite{williams1999fourier}
\begin{equation}
\nabla^2 \tilde{p}' + k_0^2 \tilde{p}'
=
\nabla\cdot\tilde{\mathbf{F}},
\label{eq:helmholtz}
\end{equation}
where $k_0=\omega/c_0$ is the acoustic wavenumber.
The solution is obtained using the free-space Green's function of the Helmholtz operator
\cite{morse1968theoretical}. Defining
$
\mathbf{r}=\mathbf{x}-\mathbf{y}$ and
$r=|\mathbf{r}|
$, 
the Green's function is
\begin{equation}
G(\mathbf{r},\omega)
=
\frac{e^{ik_0 r}}{4\pi r}.
\label{eq:greens_function}
\end{equation}
Assuming that boundary contributions vanish, the acoustic pressure can be written as
\begin{equation}
\tilde{p}'(\mathbf{x},\omega)
=
-\int_{\Omega}
\nabla_{\mathbf{r}}G(\mathbf{r},\omega)
\cdot
\tilde{\mathbf{F}}(\mathbf{y},\omega)
\,\mathrm{d}\mathbf{y},
\label{eq:greens_solution}
\end{equation}
%Here, $\mathbf{r}$ is the vector connecting the source point $\mathbf{y}$ to the observer point $\mathbf{x}$.
%
Lastly, for a harmonic point force located at the origin $\mathbf{r}=\mathbf{x}$, we have
$
\tilde{\mathbf{F}}(\mathbf{y},\omega)
=
\tilde{\mathbf{f}}(\omega)\delta(\mathbf{y}).
$
The resulting acoustic pressure field corresponding to a dipole source reads 
\begin{equation}
\tilde{p}'(\mathbf{r},\omega)
=
\frac{1}{4\pi r}
\left(
\frac{1}{r}-ik_0
\right)
\left(
\tilde{\mathbf{f}}(\omega)\cdot\hat{\mathbf{r}}
\right)
e^{ik_0 r},
\label{eq: dipole solution2}
\end{equation}
where $\hat{\mathbf{r}}=\mathbf{r}/|\mathbf{r}|$. This formulation reflects the classical dipole radiation mechanism associated with fluctuating structural forces acting on the surrounding fluid (without net mass injection) \cite{lighthill1952sound,wagner1996wind}. Under the small-amplitude vibration regime considered here, higher-order non-linear
and quadrupole volume-source contributions can be neglected. For compact sources with
a low Mach number, these terms are higher order than the surface-loading dipole
term, which therefore provides a leading-order description of the radiated acoustic
field \cite{lighthill1952sound,Curle1955,FfowcsWilliamsHawkings1969}.

% ############################################################################# %
\section{Radiation‑Impedance Correction Coefficient $\gamma$}
\label{ap: gamma}

The equivalent dipole model described in Section~\ref{sec:rad} represents the submerged structure as a discrete set of point dipole sources. For a structural node \(n\) with effective mass \(m_{\mathrm{eff},n}\) and complex acceleration \(\mathbf{a}_n(\omega)\) (time convention \(\mathrm{e}^{-\mathrm{i}\omega t}\)), the force acting on the fluid is locally proportional to the inertial force. However, the finite cross section of the structural members modifies the acoustic radiation with respect to that of an ideal line of compact dipoles. This appendix derives the frequency‐-dependent correction coefficient \(\gamma(\omega, R_n)\) introduced in Eq.~\ref{eq:equivalent_force}, which accounts for the radiation impedance of a cylindrical radiator.

The derivation proceeds by comparing the exact surface pressure of a harmonically oscillating rigid cylinder with the pressure field produced by a line distribution of dipoles of equivalent strength, and then identifying the factor \(\gamma\) that reconciles both descriptions.

% ---------------------------------------------------------------------------- %
\subsubsection*{Exact pressure on a rigid oscillating cylinder}

Consider an infinitely long rigid cylinder of radius \(a\) immersed in an ideal fluid of density \(\rho\) and sound speed \(c\). The cylinder oscillates harmonically in the \(x\)-direction with velocity amplitude \(V_a\). For the \(\mathrm{e}^{-\mathrm{i}\omega t}\) convention adopted throughout this work, outgoing cylindrical waves are represented by Hankel functions of the first kind. The acoustic pressure on the cylinder surface \(r=a\), as a function of the azimuthal angle \(\theta\), is therefore given by the classical solution~\cite{morse1968theoretical,Fahy2007Cap4}
\begin{equation}
    p_{\mathrm{cyl}}(a,\theta,\omega) = \mathrm{i}\rho c\, V_a \frac{H_1^{(1)}(ka)}{H_1^{(1)\prime}(ka)} \cos\theta,
    \label{eq:p_cyl}
\end{equation}
where \(k=\omega/c\) is the acoustic wavenumber, \(H_1^{(1)}\) is the Hankel function of the first kind and order one, and \(H_1^{(1)\prime}\) denotes its derivative with respect to the argument. The time factor \(\mathrm{e}^{-\mathrm{i}\omega t}\) is omitted throughout. Equation~\eqref{eq:p_cyl} describes a dipole ‐-like radiation pattern that fully incorporates fluid loading and the finite cross‑section effects.

% ---------------------------------------------------------------------------- %
\subsubsection*{Pressure radiated by a line dipole}

The uncorrected equivalent source model replaces the cylindrical member by a continuous line of dipole sources of strength per unit length \(F\) (orientated along the \(x\)-axis). The free‑space Green’s function of the Helmholtz equation in three dimensions is
\begin{equation}
    G_{3D}(R) = \frac{\mathrm{e}^{\mathrm{i}kR}}{4\pi R}, \qquad R = \sqrt{r^{2}+(z-z')^{2}},
\end{equation}
with \(r = \sqrt{x^{2}+y^{2}}\). For an observer located in the plane \(z=0\) at \((r,\theta)\), the pressure field generated by a uniform line dipole that extends along the \(z\)-axis is
\begin{equation}
    p_{\mathrm{dip}}(r,\theta,\omega) = \int_{-\infty}^{\infty} F\,\frac{\partial G_{3D}}{\partial x}\,\mathrm{d}z'
      = \frac{F}{4\pi} \frac{\partial}{\partial x} \int_{-\infty}^{\infty} \frac{\mathrm{e}^{\mathrm{i}kR}}{R}\,\mathrm{d}z'.
    \label{eq:p_dip_int}
\end{equation}
The integral is evaluated at \(\mathrm{i}\pi H_0^{(1)}(kr)\)~\cite{jensen2011computational}. Performing differentiation with respect to \(x = r\cos\theta\) and using \(H_0^{(1)\prime}(\xi) = -H_1^{(1)}(\xi)\) yields
\begin{equation}
    p_{\mathrm{dip}}(r,\theta,\omega) = -\frac{\mathrm{i}kF}{4} \cos\theta\, H_1^{(1)}(kr).
    \label{eq:p_dip}
\end{equation}
Equation~\eqref{eq:p_dip} represents the radiation from a line of ideal dipoles in an unbounded fluid.

% ---------------------------------------------------------------------------- %
\subsubsection*{Determination of \(\gamma\)}

In the discrete model, the uncorrected force per unit length is taken as the inertial force of the structural member:
\begin{equation}
    F_0 = m' A,
\end{equation}
where \(m' \approx \rho\pi a^{2}\) is the mass per unit length of the cylinder and \(A = \mathrm{i}\omega V_a\) is the complex acceleration amplitude. It should be noted that the structural mass has been neglected under the assumption of a thin wall structure, i.e., \(m_{\text{s}}=2\rho_s\pi a t\ll\rho\pi^2 a\), where t is the thickness of the wall. The uncorrected line‑dipole pressure would then be \(p_{\mathrm{dip},0} = p_{\mathrm{dip}}(F_0)\).

The correction coefficient \(\gamma\) is defined such that the exact surface pressure \(p_{\mathrm{cyl}}\) is recovered when the dipole strength is scaled by \(\gamma\), i.e.,\ \(p_{\mathrm{cyl}} = p_{\mathrm{dip}}(\gamma F_0)\). Because \(p_{\mathrm{dip}}\) is linear in \(F\),
\begin{equation}
    \gamma(\omega,a) = \frac{p_{\mathrm{cyl}}(a,\theta,\omega)}{p_{\mathrm{dip}}(a,\theta,\omega)\big|_{F=F_0}}.
    \label{eq:gamma_def}
\end{equation}
Substituting~\eqref{eq:p_cyl} and~\eqref{eq:p_dip} and cancelling the common factor \(\cos\theta\) yields the following.
\begin{equation}
    \gamma = \frac{\displaystyle \mathrm{i}\rho c V_a \frac{H_1^{(1)}(ka)}{H_1^{(1)\prime}(ka)}}
               {\displaystyle -\frac{\mathrm{i}k}{4} (m'A) H_1^{(1)}(ka)}.
\end{equation}
Using \(A = \mathrm{i}\omega V_a\), \(m' = \rho\pi a^{2}\) and \(k = \omega/c\), the Hankel functions cancel, and we obtain the following
\begin{equation}
    \gamma(ka) = \frac{4\mathrm{i}}{\pi (a k)^{2} H_1^{(1)\prime}(ka)}.
    \label{eq: gamma final}
\end{equation}
Equation~\eqref{eq: gamma final} is the radiation correction coefficient impedance ‐ applied in Eq.~\ref{eq:equivalent_force} of the main text, where \(a\) is the external radius \(R_n\) of the cylindrical member. For rectangular cross sections ‐ an equivalent area‑preserving radius is used. The coefficient is complex, capturing both amplitude and phase modifications as a result of the distributed nature of the radiator.

\end{document}